# Secure Debit Card Device Model

By

| | |
|---|---|
| **Rumaisah Munir** | **(2009-NUST-BE-BICSE-202)** |
| **Saad Bin Khalid** | **(2009-NUST-BE-BEE-311)** |

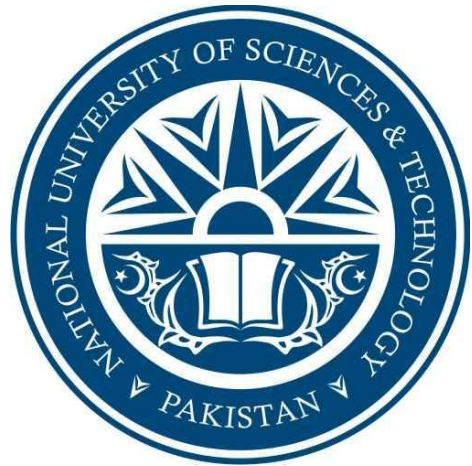

**Submitted to NUST SEECS in fulfillment of the requirement for the award of the degree of Bachelors of Electrical Engineering**

**NUST School of Electrical Engineering & Computer Science National University of Sciences & Technology**

**Islamabad, Pakistan**

**2013**

1 | P a g e

# CERTIFICATE

It is certified that the contents and form of the Final year project report entitled **'Secure Debit Device Model'** submitted by Rumaisah Munir (2009-NUST-BE-BICSE-202) and Saad Bin Khalid (2009-NUST-BEE-315) have been found to be satisfactory for the requirement of the degree.

**Advisor: Dr. Abdul Ghafoor Abbasi**

**Co-Advisor: Dr. Awais Kamboh**

**Co-Advisor: Mr. Nasir Mahmood**

# DEDICATION

To our Mentors

&

To our Parents

&

To our Colleagues

# ACKNOWLEDGEMENTS

First and foremost, we would like to express our sincere gratitude to our advisor Dr. Abdul Ghafoor for the continuous support of design and development of our Final Year Project, for his patience, motivation, enthusiasm, and immense knowledge. His guidance helped us in all the time of research and completion of this project. We could not have imagined having a better advisor and mentor for our Final Year Project.

Besides our advisor, we are heartily thankful to our co-advisor, Dr. Awais Kamboh, whose encouragement, guidance and support from the initial to the final level enabled us to develop an understanding of the project.

Our sincere thanks also go to Mr. Nasir Mahmood for offering us the significant assistance during development phase of this project.

Lastly, we offer our regards and blessings to all of those who supported us in any respect during the completion of the project.

# Contents

# Table of Figures

## Tables:

# ABSTRACT

The project envisages the implementation of an e-payment system utilizing FIPS-201 Smart Card. The system combines hardware and software modules. The hardware module takes data insertions ( e.g. currency notes), processes the data and then creates connection with the smart card using serial/USB ports to perform further mathematical manipulations. The hardware interacts with servers at the back for authentication and identification of users and for data storage pertaining to a particular user. The software module manages database, handles identities, provide authentication and secure communication between the various system components. It will also provide a component to the end users. This component can be in the form of software for computer or executable binaries for PoS devices.

The idea is to receive data in the embedded system from data reader and smart card. After manipulations, the updated data is imprinted on smart card memory and also updated in the back end servers maintaining database. The information to be sent to a server is sent through a PoS device which has multiple transfer mediums involving wired and unwired mediums. The user device also acts as an updater; therefore, whenever the smart card is inserted by user, it is automatically updated by synchronizing with backend database.

The project requires expertise in embedded systems, networks, java and C++ (Optional).

# CHAPTER 1

# Introduction To Project

The Significance of "Electronic Purse" is highly diverse, for security reasons, as it seems inevitable. The world seems to be replacing all the current financial systems to smart card based payment systems, from banks to toll stations. To cater for the need of a highly efficient secure e-payment system, we have implemented proficient and effective smart card based e-payment prototype to perform four different sorts of transactions.

The Significance of Smart Card Technology is evident in SIM Card Technology, Telecommunications, for securing digital content and physical assets, in healthcare informatics, physical access, enterprise and network security.

First part of Chapter 1 describes the basic concepts of smart card system, most common terms used in smart card based systems. Second part of this chapter discusses how smart card security works. The third part discusses the components of smart card system and main modules. The fourth part elaborates the transaction theory. The last part of this chapter involves the transaction theory and mathematical modeling of main blocks of a smart card based security system. Chapter 3 includes the literature review while the chapter 4 explains the complete methodology of the project. This chapter includes the hardware architecture and the software architecture of the project. Chapter 5 includes the results.

## *1.1 Preliminaries:*

This section deals with fundamental concepts of Smart Card Security, types of Card Systems and components of a Smart Card.

### 1.1.1 Smart Card Basics:

A smart card is a minute-sized card consisting of embedded ICs (Integrated Circuits). It is also called an "embedded IC card". Smart Cards include a microprocessor in itself and a supporting, operating software. The material of smart cards is plastic or PVC. Smart cards do not need an internal power source for functioning. Cryptographic protocols are used for secure communication between smart card and other devices.

There are two types of smart cards:

- Contact smart cards.
- Contactless smart cards.

### 1.1.1. Contact smart cards:

These are inserted in the smart card reader. The processor in the smart card interacts with the reader and system of the reader. Power is supplied to the smart card, by the smart card reader. A communication channel is established between the smart card reader and the smart card when the smart card is inserted in the reader.

### 1.1.2. Contactless smart cards:

These type of cards make use of short-range RF communications (at data rates of 106–848 kilobit/s). which helps in the commencement of a wireless connection. Such a mechanism takes place when the contactless smart card is brought near to a contactless terminal. The technology is known as RF Induction technology. The card only requires the proximity of an antenna. An inductor is present in such a smart card which captures the RF signal, rectifies it to a reasonable extent and powers up the electronic components of the card.

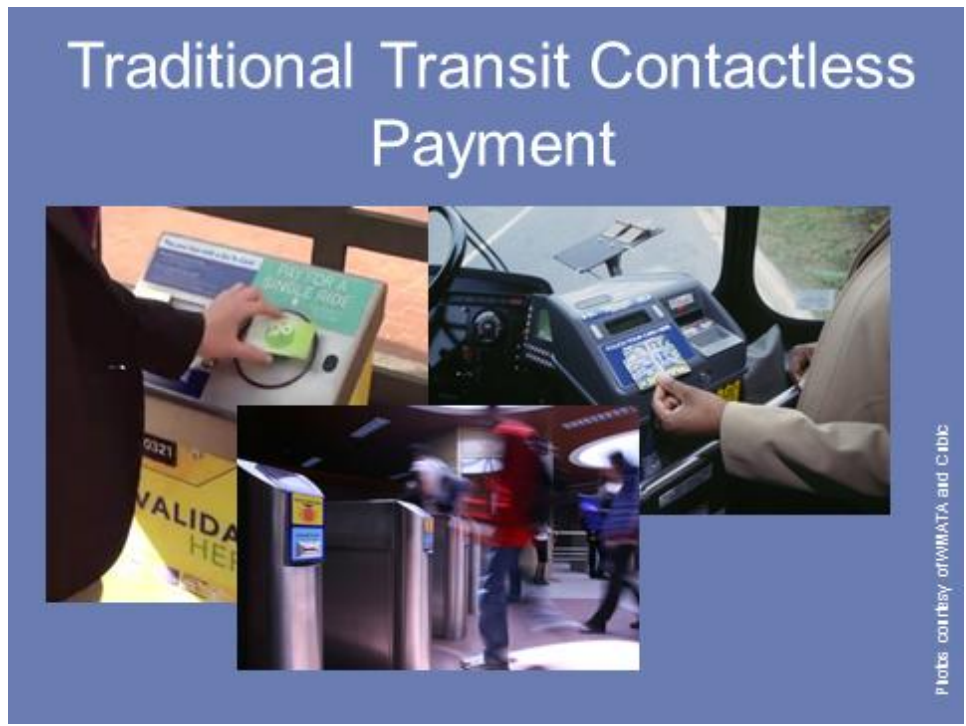

**Figure 1.1: Application of Contactless Smart Cards (Taken from [2])**

### 1.1.3. Smart Card Specifications:

A normal smart card has three separate memory stores. Two from among these are ROMs (read only memory) and one is RAM (random access memory) - 8 kilobytes of RAM, 346 kilobytes of ROM and an additional programmable ROM with 256 kilobytes of memory, controlled through a 16-bit microprocessor.

### 1.1.4. Functions:

- User Identification.
- User Authentication.
- Storage of Data.
- Processing of applications.

### 1.1.5. Application Areas:

- "Electronic purse" is based on a smart card system. Funds are stored on the card. No network connectivity is required.
- Smart cards are common all over the world as mobile SIMs are presumably smart cards.
- Smart Cards provide very high security.
- Smart cards are used as "electronic wallets". Smart card is loaded with cash amount.
- Smart-cards are used for identity authentication. For this purpose a PKI (Public Key Infrastructure) is used. The smart card is capable of storing digital certificate and other information provided with the PKI.

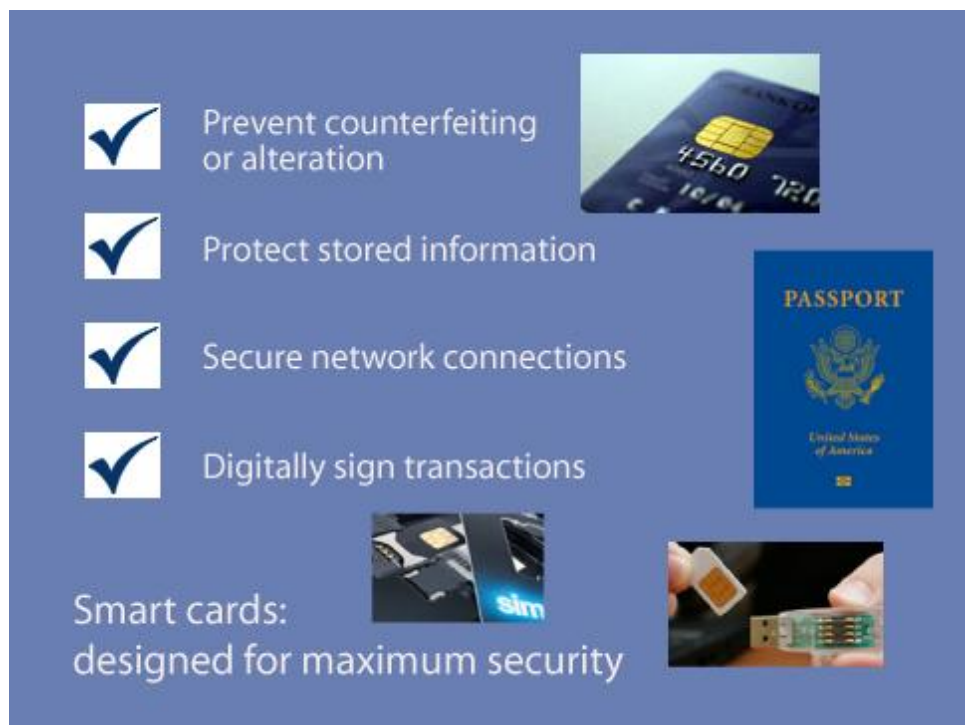

**Figure 1.2: Security provided by Smart Cards (Taken from [2])**

### 1.1.6. How Smart Card Security Works?

The microprocessor that is built inside a smart card, has includes either public key certificates or other various types of cryptographic protocols. Digital identity credentials are also present which assists in authenticating and authorizing for accessing the systems. The smart card works on the principle of randomness. A smart card's cryptographic protocols help in generating a random number from input data and stored secret keys. The random number is sent back to server. The server is responsible for verification of the number, which if verified, indicates that the card is present as well as authentic.

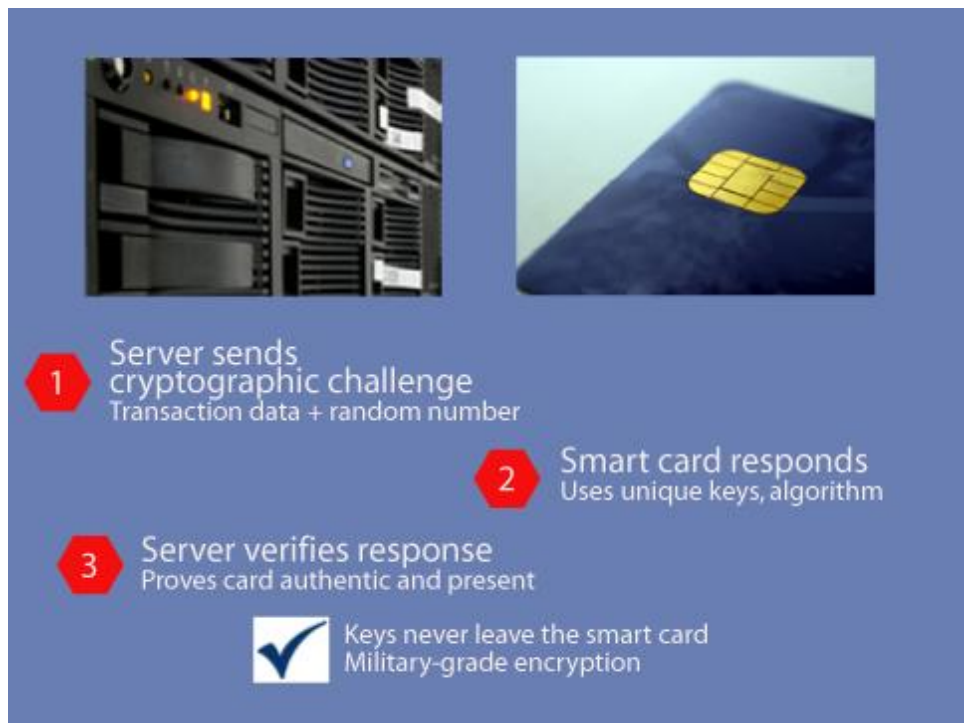

**Figure 1.3: Server authenticates Smart Cards (Taken from [2])**

### 1.1.7. Mutual Authentication:

Smart cards are capable of authenticating the server. The process is called mutual authentication. When the smart card is utilised for authenticating the server, relay attacks are prevented. In mutual authentication, a server is authenticating the smart card and vice versa.

### 1.1.8. Two-Factor Authentication:

In such an authentication mechanism, the server authenticates an account on the basis of smart card, besides username and password. This provides ultra security, because even if the username and password are hacked, a person will not be able to login without smart card.

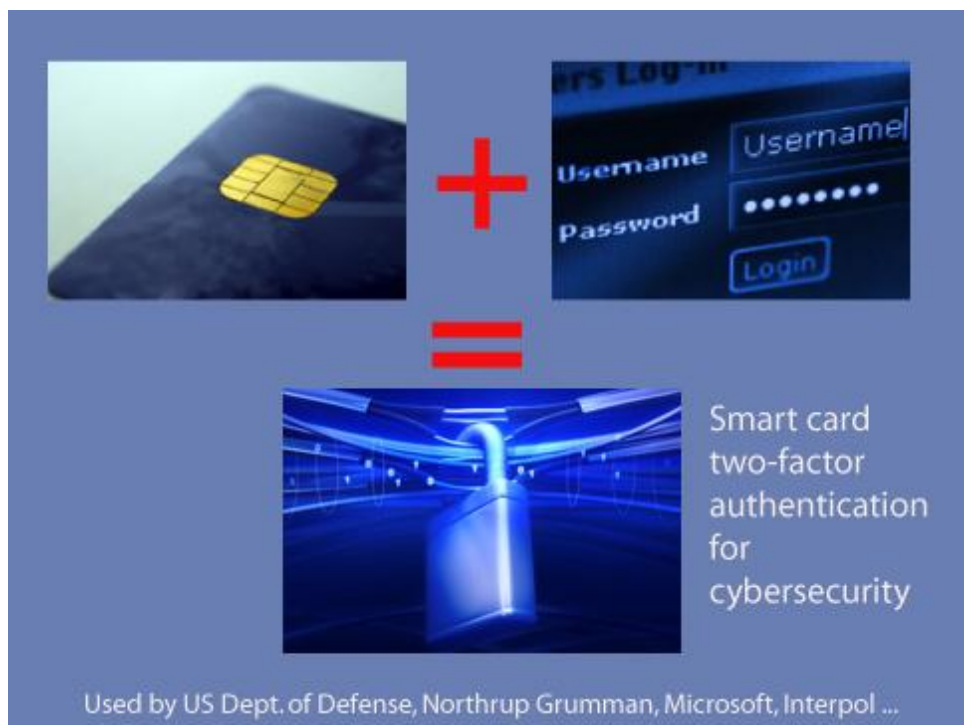

**Figure 1.4: Two-Factor Authentication (Taken from [2])**

### 1.1.9. Transaction Theory:

The transaction processing requirements are very time consuming. They require receipts and signatures. Because of the time consumption, there is a switch from traditional manual systems to contactless transaction systems. There is no direct physical connection between the POS (Point-Of-Sale) terminal and the user's device (card). This removes the hassle of time consumption.

Benefits of such a system are:

- Improved customer experience.
- Reduced payment media issuance.
- Interagency interoperability.
- Increased revenue opportunities.

## *1.2 Problem Statement And Expected Results:*

In a third world country like Pakistan, financial frauds are easy to occur and hard to counter. In recent years, billions of frauds have been avoided in the developed countries. Financial security is a need in a third world country like Pakistan. A system needs to be designed which will ensure data security and financial security. We create a prototype for a financial system, based on smart card security

## *1.2 Requirement Analysis:*

**1.1.10. Server Side:**

The software will be hosted on the hardware device connected to the backend server and SQL 2008 Database server. The web server is listening to the port 8080.

**1.1.11. Client Side:**

The system is a web based application; clients are required to have a modern web browser such as Mozilla Firefox 3.0, Internet Explorer 8 and enable cookies. Internet connection is most needed to access this system.

# Chapter 2

# Literature Review

## *2.1 Secure Architecture:*

The smart card has a secure architecture. The BIOS (security kernel in a smart card) concept is the interaction layer between the hardware and the software. The security kernel features of a smart card kernel are as follows:

- None of the applications may take the control of the core processor.
- Enforcement of mapping of data between data storage and applications.
- Every application has to provide for and operate it's own security protocols.

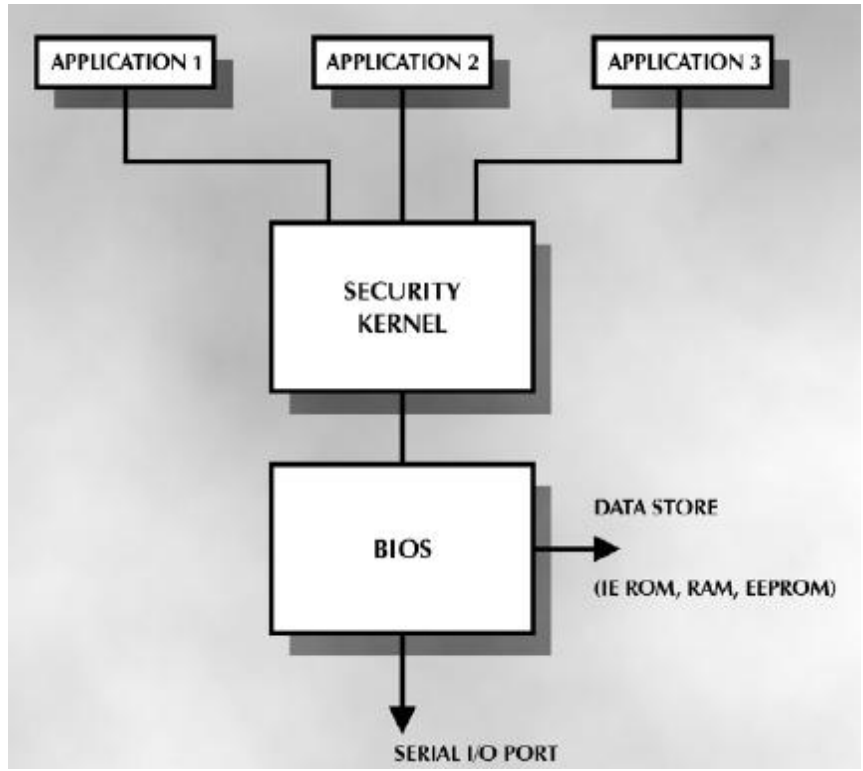

**Figure 2.1: Security Architecture of a Smart Card (Taken from [4])**

## *2.2. Multiple Applications Strategy:*

Smart cards are the way forward when it comes to interoperability. Smart cards are being fabricated to be able to support multiple applications. By multiple applications, it is meant that a multitude of applications are supported on the card. This removes the hassle of holding other cards. Adding more applications to the card ensures adding of more value.

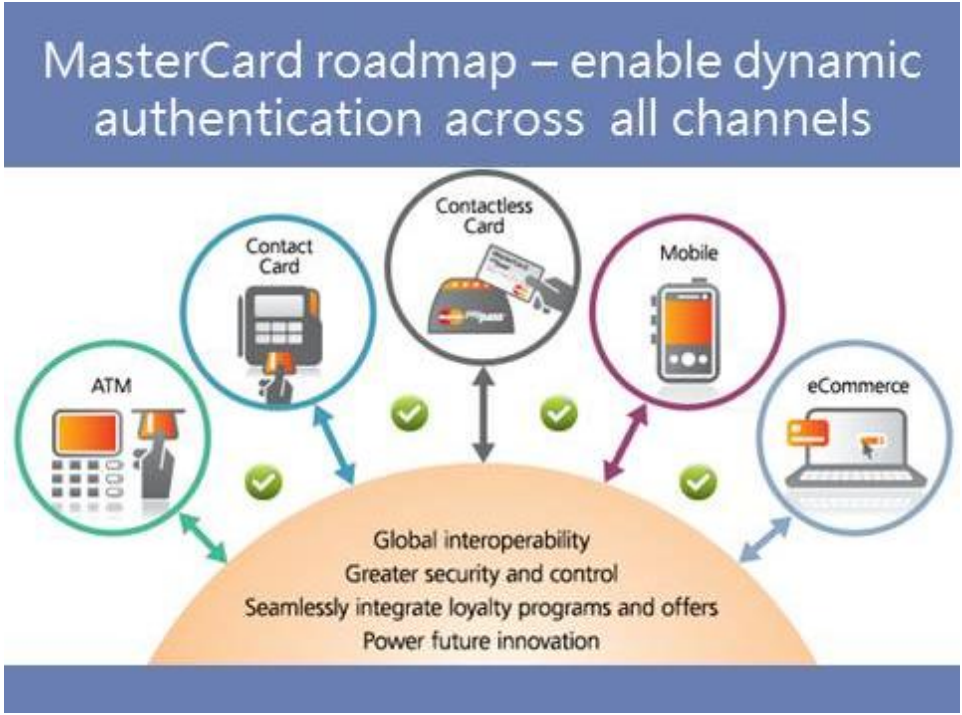

**Figure 2.2: Multiple Application Support (Taken from [1])**

## *2.3. Magnetic Stripe VS Smart Cards:*

| Requirement | Smart Card | Magnetic Stripe Cards |
|---|---|---|
| Fraud reduction | 🔵 | 🔴 |
| Accurate individual identification | 🔵 | 🔴 |
| Accurate information | 🔵 | 🔴 |
| Card re-issuance cost elimination | 🔵 | 🔴 |
| Secure and authenticated access | 🔵 | Feature not available |
| Information exchange support | 🔵 | 🔴 |

| Feature | Smart Card | Magnetic Stripe Card |
|---|---|---|
| Privacy and security improvement | 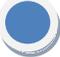 | Feature not available |
| Real-time portable electronic record | 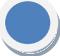 | Feature not available |
| Immediate emergency access to data | 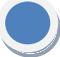 | Feature not available |
| Two factor authentication for accessing records online | 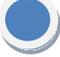 | Feature not available |
| Digital signature support for enabling strong audit | 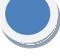 | Feature not available |
| Storage Capacity | 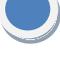 | 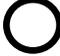 |
| Security Features | 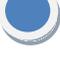 | 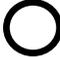 |
| Modification after issuance | 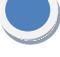 | 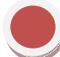 |
| Cost of ID Device | 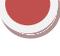 | 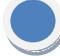 |
| Cost of reader | 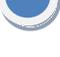 | 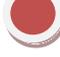 |
| Multiple Application support | 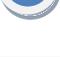 | Feature not available |
| On-card biometric storage | 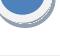 | Feature not available |
| On-card biometric match | 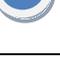 | Feature not available |

**Table 2.1: Comparison of Smart Card with Magnetic Stripe Card (Taken from [3])**

**Legend: Best**      **Average**      **Worst**

# CHAPTER 3

# Methodology

This chapter is divided into four parts. The first part shows the flow charts for the project. Second part shows the Use Case diagrams. The third part shows the architecture diagram. And the last part consists of the generic security commands based on JSON format used for transmission of data between server and client.

## *3.1. Functions:*

There are four functions being implemented in our prototype:
- Account To Account Transfer.
- Pay Over The Counter.
- Cash Deposit.
- Cash Withdrawal.

## *3.2. Flow Charts*

### 3.2.1. Account To Account Transfer

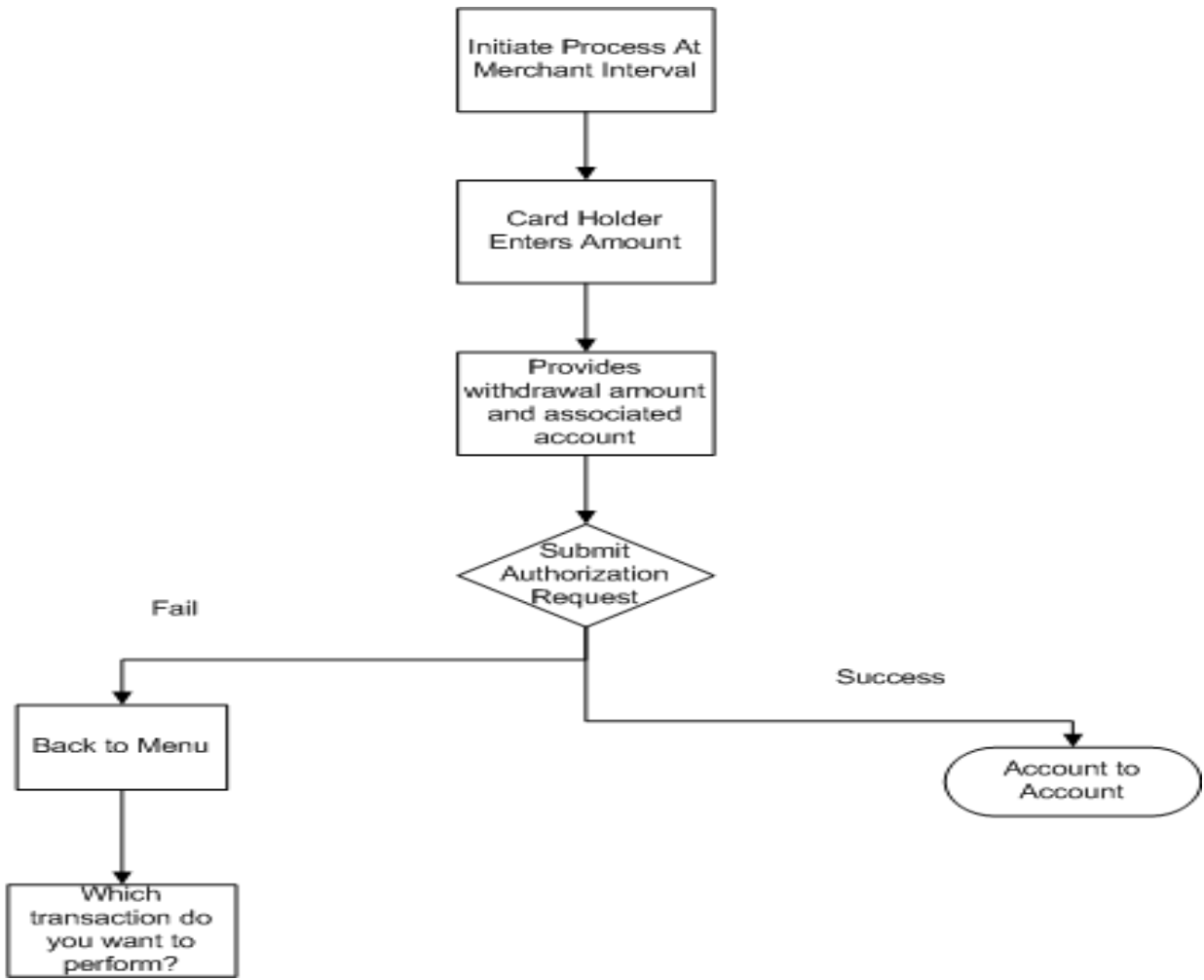

**Figure 3.1: Flow Chart for Account To Account Transaction**

**3.2.2. Pay Over The Counter**

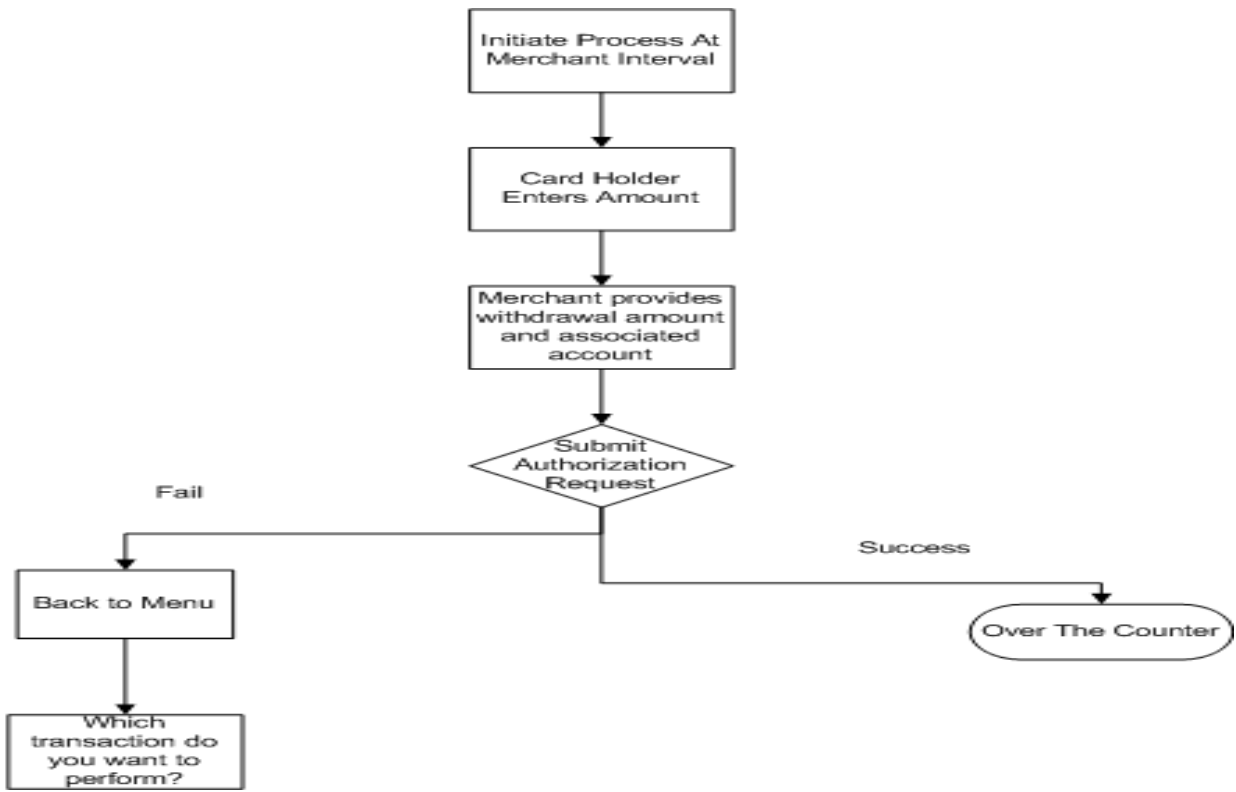

**Figure 3.2: Flow Chart for Pay Over The Counter Transaction**

**3.2.3. Cash Deposit**

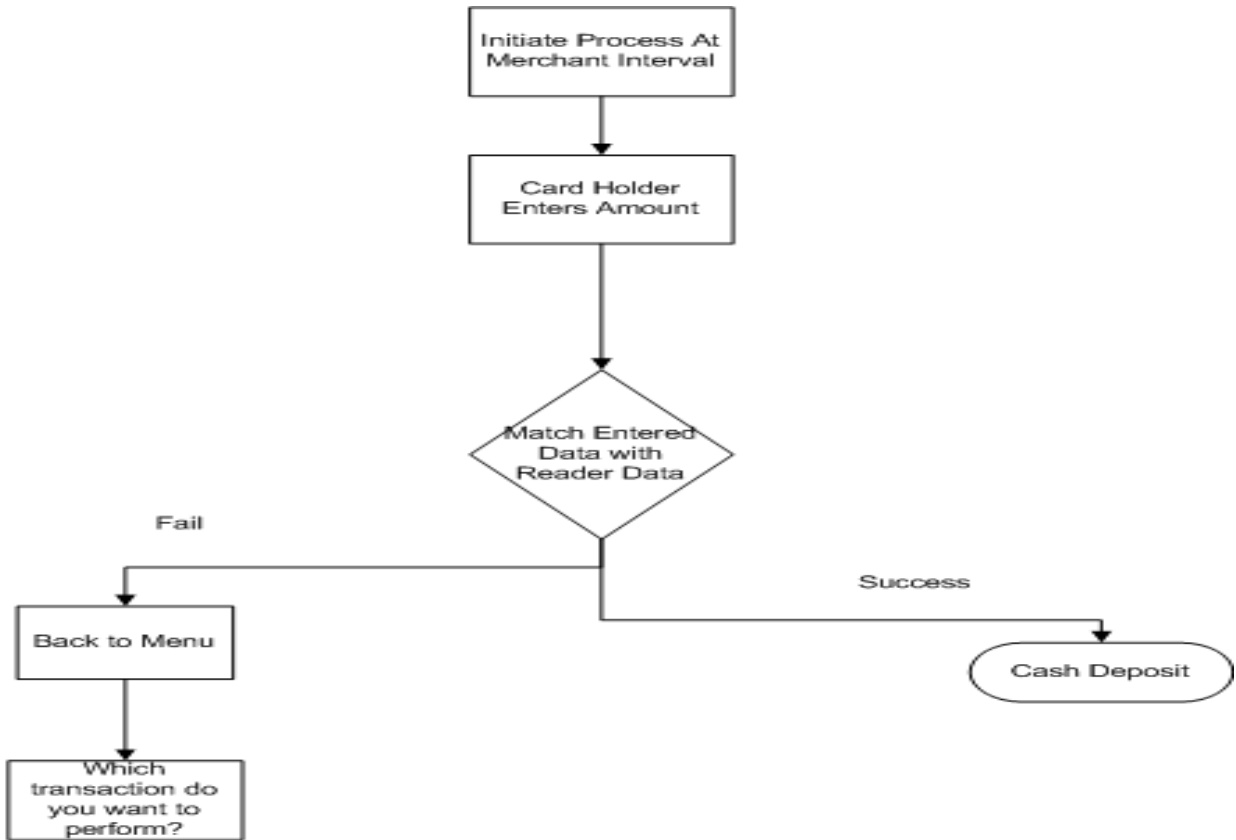

**Figure 3.3: Flow Chart for Cash Deposit Transaction**

## 3.2.4. Cash Withdrawal

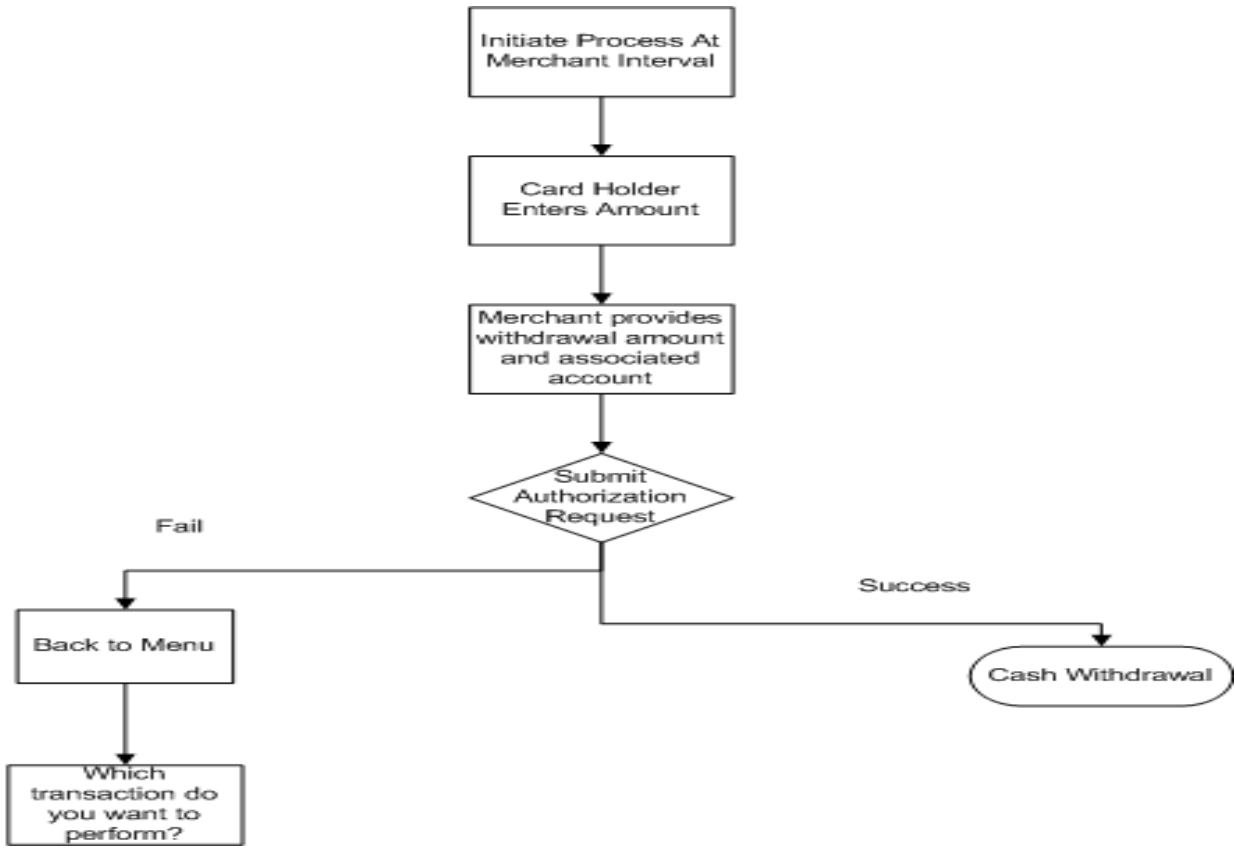

**Figure 3.4: Flow Chart for Cash Withdraw Transaction**

## 3.2.5. System Flow Chart

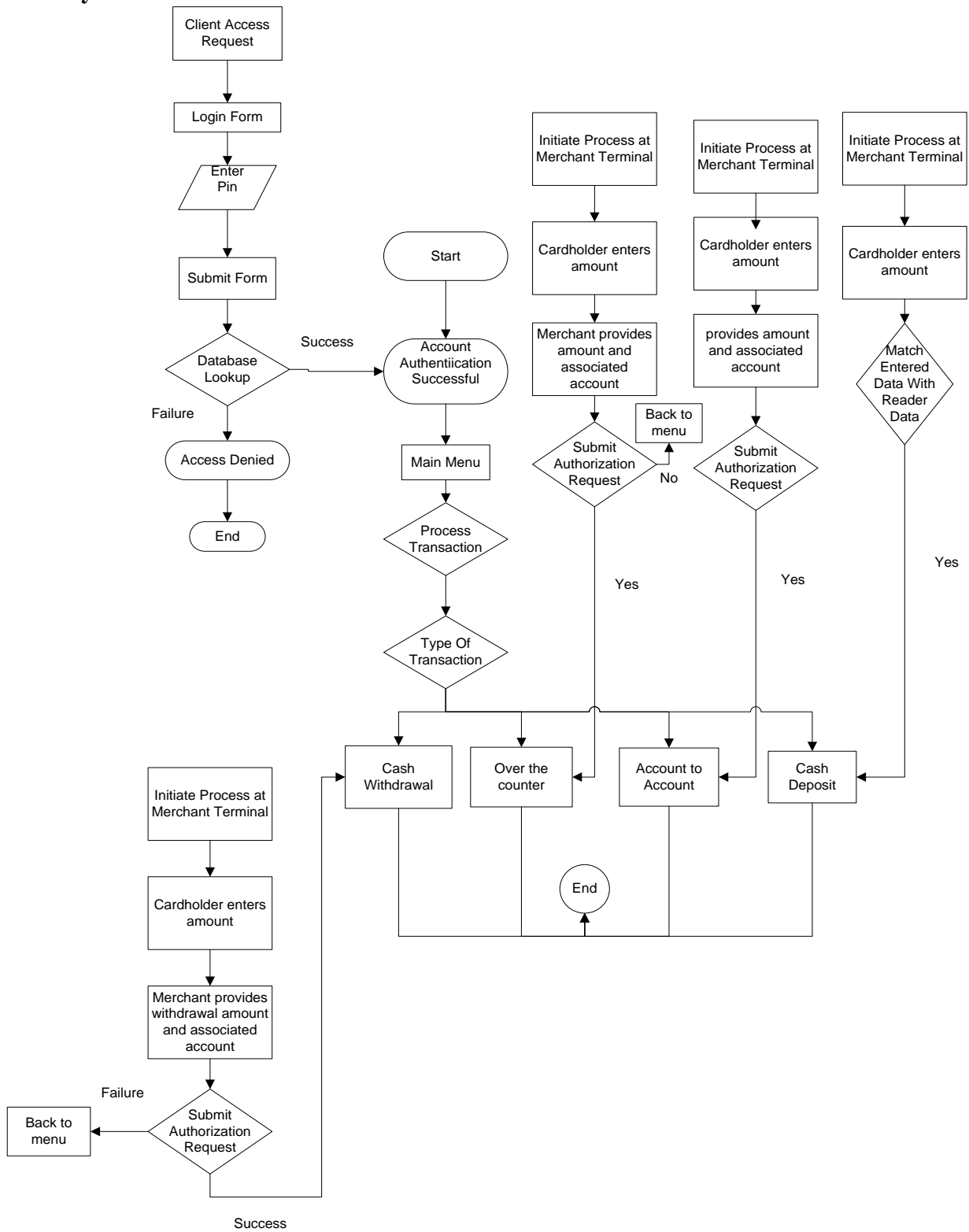

**Figure 3.5: System Flow Chart**

## *3.3. Use Cases:*
### 3.3.1. User-Client Use Case:

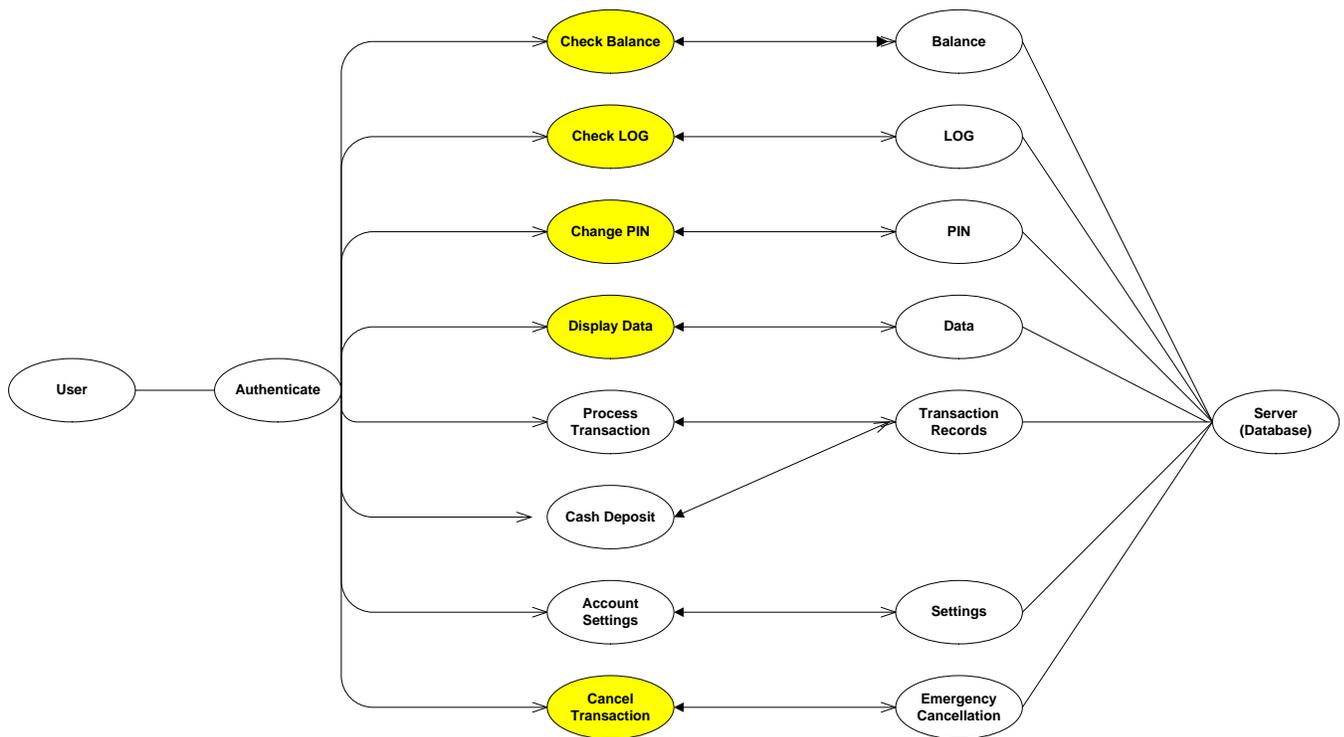

**Figure 3.6: User-Client Use Case Model**

## 3.3.2. Admin-User Use Case:

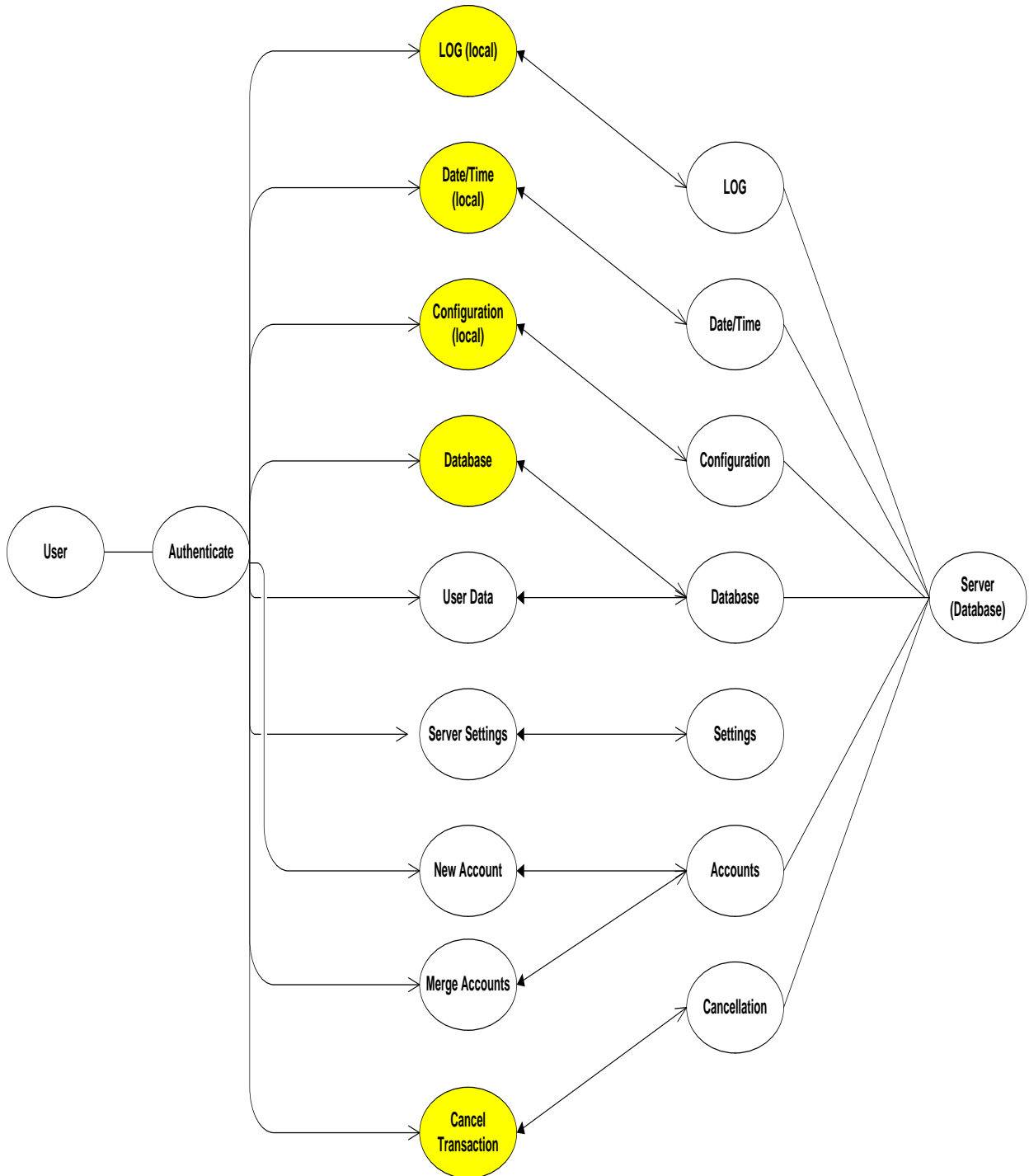

**Figure 3.7:Admin-User Use Case**

### 3.3.3. Simple Client Use Case:

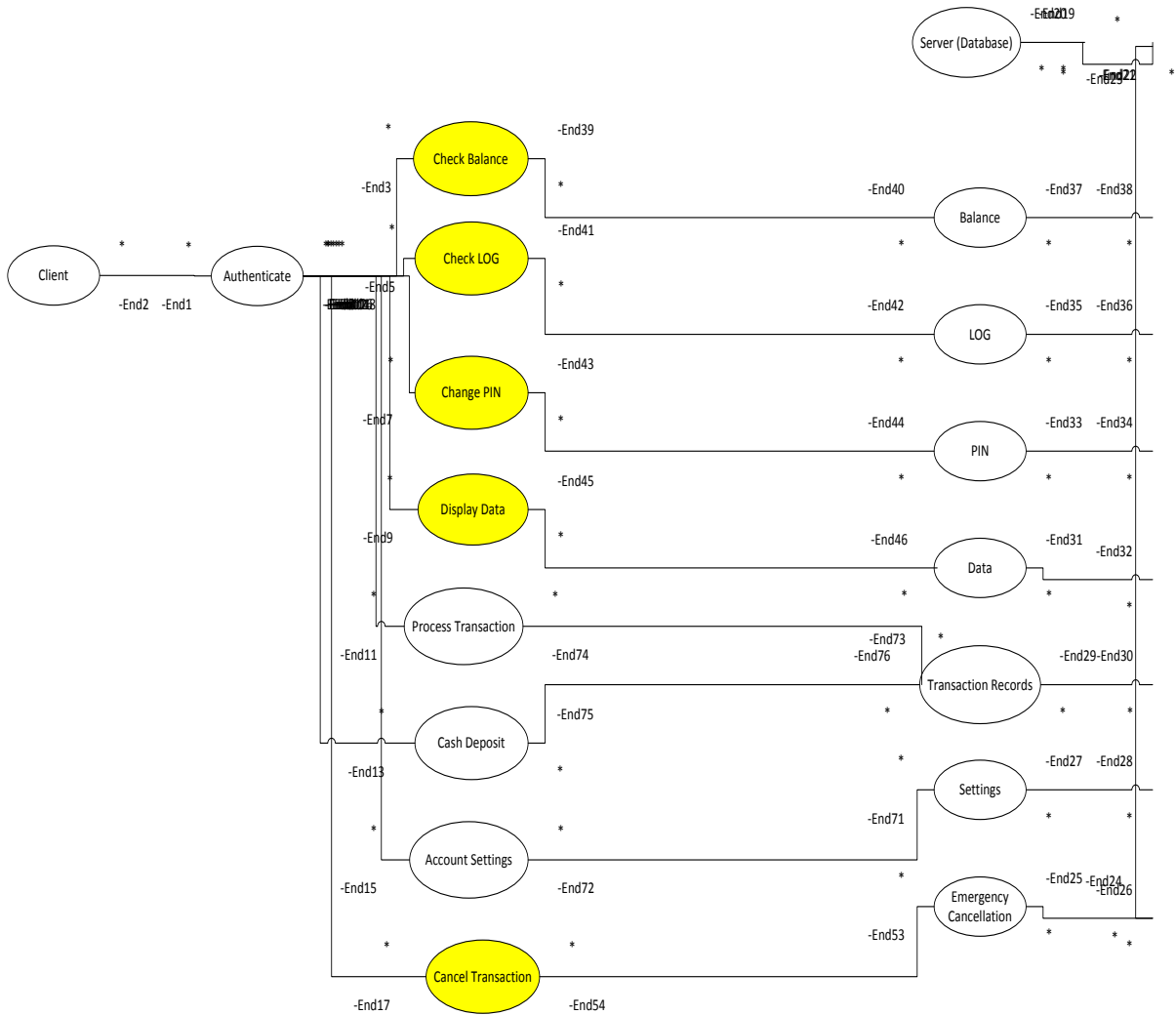

**Figure 3.8: Simple Client Use Case Model**

## 3.3.4. Admin Client Use Case:

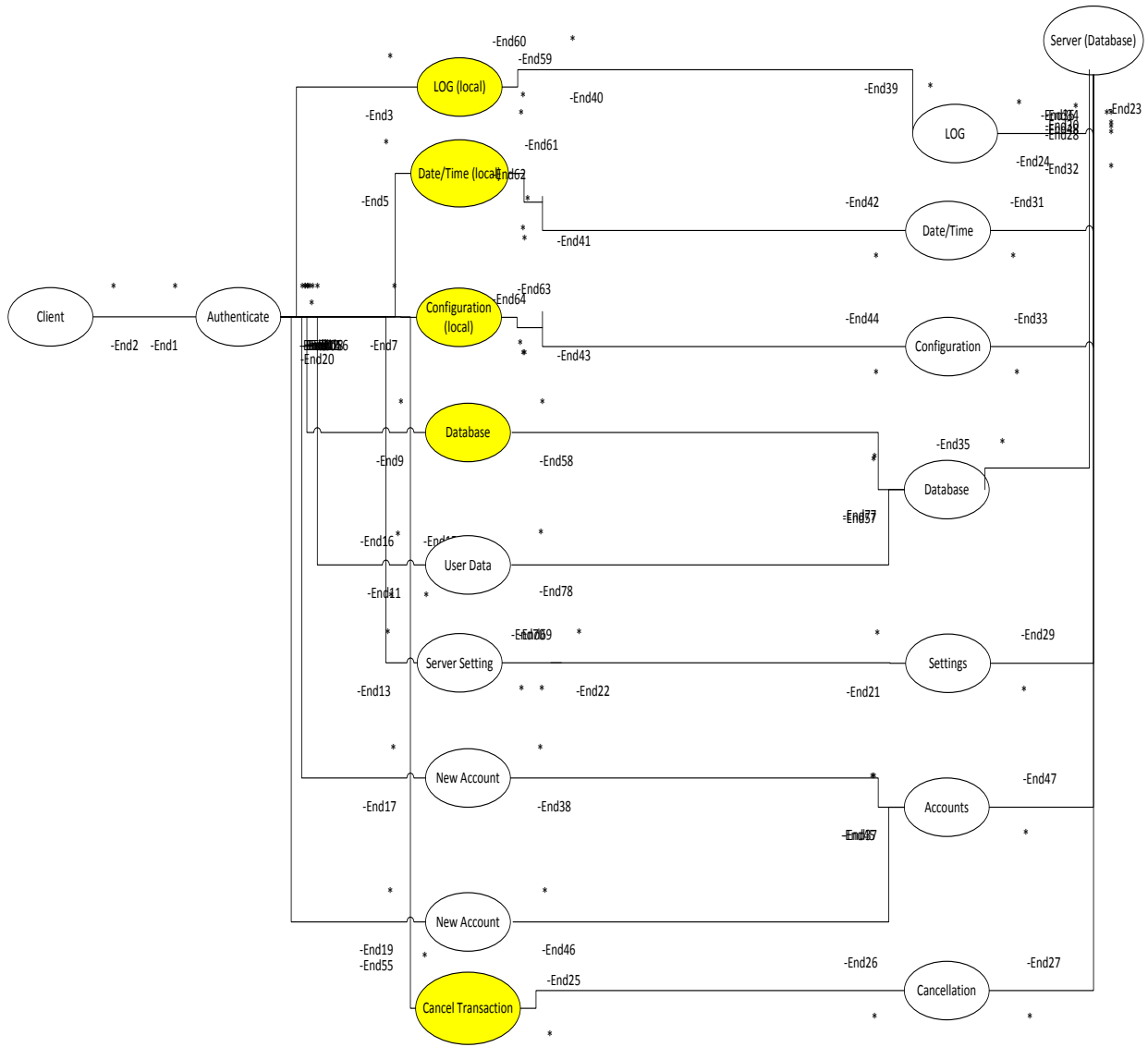

**Figure 3.9: Admin-Client Use Case**

## 3.4. Architecture Diagram:

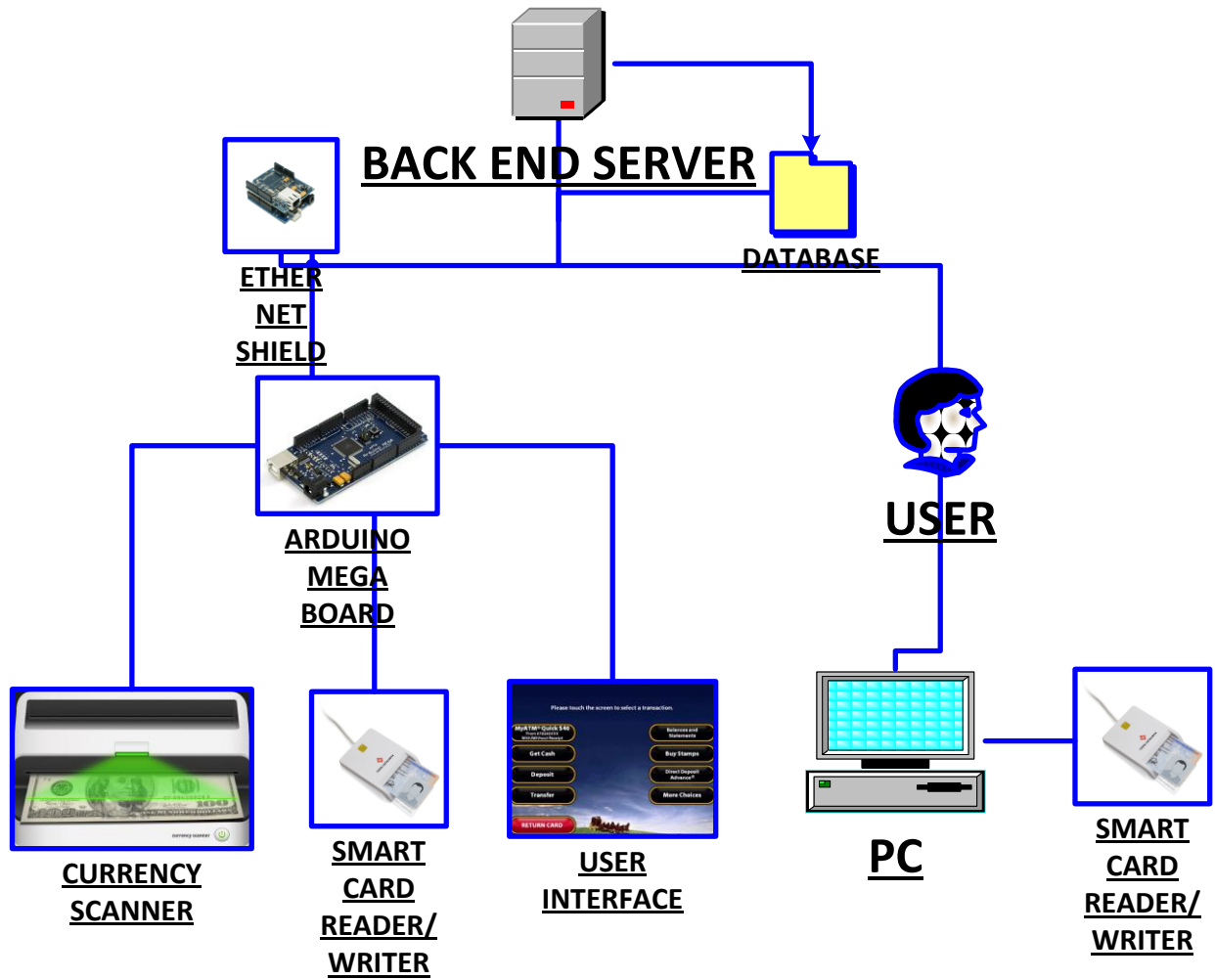

Figure 3.10: Architecture Diagram

## *3.5. Entity-Relationship Diagram:*

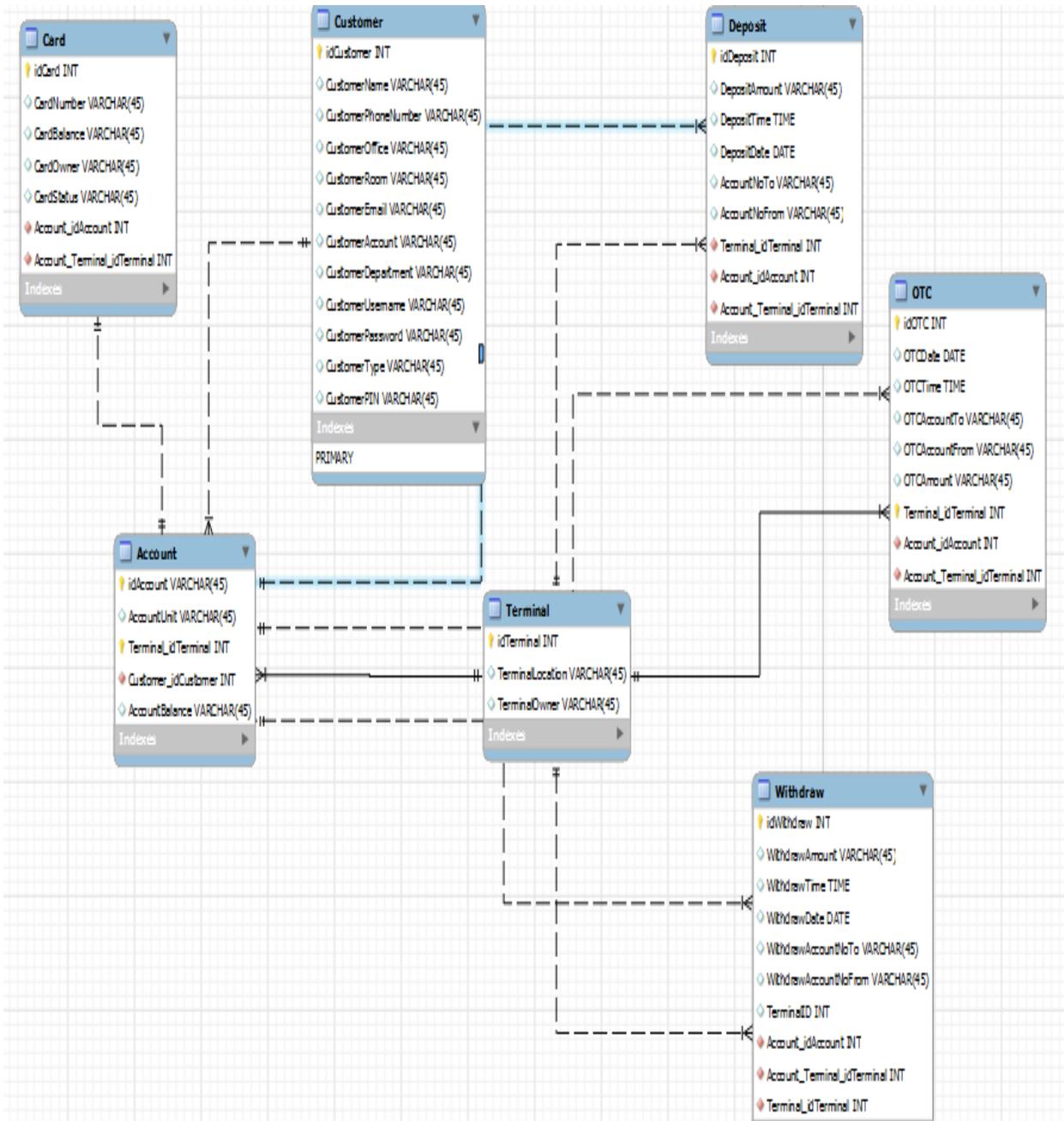

**Figure 3.11: Entity Relationship Diagram**

## *3.6. JSON Commands:*

**3.6.1. Pay Over The Counter:**

**3.6.1.1. Enter amount message:**
{
 "Method": "EnterAmount"
 "To Account": "Merchant"
 "From Account": "User"
 "Amount": " 100"
 "Time": "1100 hours"
 "Date": 13-5-2012
 }

**3.6.1.2. Response message:**
{
 "Method": "EnterAmount"
 "Result": "OK",
"To Account": "Merchant",
 "From Account": "User",
 "Amount": " 100",
 "Time": "1100 hours",
 "Date": "13-5-2012"
"Error": "null"
 }

**3.6.1.3. Enter PIN message:**
{

"Method": "EnterPIN"
"Message": "Please insert PIN:"
"Time": "1100 hours"
"Date": 13-5-2012
}

### 3.6.1.4. Response message:
{
"Method": "EnterPIN"
"Result": "PIN",
"Time": 1100 hours
"Date": 13-5-2012"
"Error": "null"
}

### 3.6.1.5. Verification message:
{
"Message": "VerifyPIN",
"Method": "PIN",
"Time": "1100 hours",
"Date": "13-5-2012"
}

### 3.6.1.6. Response message#1:
{
"Message Type": "VerifyPIN"
"Result": "Verified",
"Time": "1100 hours",
"Date": "13-5-2012"
"Error": "null"

}

### 3.6.1.7. Response message#2:
{
 "Message Type": "VerifyPIN"
 "Result": "NotVerified",
"Time": "1100 hours",
 "Date": "13-5-2012"
"Error": "Verification Unsuccessful"
 }

### 3.6.1.8. Transmission message:
{
 "Method": "Transmit"
 "To Account": "Merchant"
 "From Account": "User"
 "Amount": "Rs. 100"
 "Time": "1100 hours"
 "Date": 13-5-2012
 }

### 3.6.1.9. Response message#1:
{
 "Method": "Transmit"
 "Result": "OK",
"To Account": "Merchant",
 "From Account": "User",
 "Amount": "Rs. 100",
 "Time": "1100 hours",
 "Date": "13-5-2012"

"Error": "null"
}

### 3.6.1.10. Response message#2:
{
 "Method": "Transmit"
 "Result": "NotTransmitted",
"To Account": "Merchant",
 "From Account": "User",
 "Amount": "Rs. 100",
 "Time": "1100 hours",
 "Date": "13-5-2012"
"Error": "Account Not Found"
}

### 3.6.1.11. Response message#3:
{
 "Method": "Transmit"
 "Result": "NotTransmitted",
"To Account": "Merchant",
 "From Account": "User",
 "Amount": "Rs. 100",
 "Time": "1100 hours",
 "Date": "13-5-2012",
"Error": "Account Has Not Enough Cash"
}

### 3.6.2. Transaction Account To Account

### 3.6.2.1. Enter amount message:

{
 "method": "Amount"
 "To Account": "User No.1"
 "From Account": " User No.2"
 "Time":  "1100 hours"
 "Date": "13-5-2012" }

**3.6.2.2. Response:**
{
"Method": "EnterAmount",
 "Result" :  "Okay",
 "To Account": User No.1
 "From Account": User No.2
 "Error": "null",
 "Time": 1100 hours
 "Date": 13-5-2012
}

**3.6.2.3. Enter PIN message:**
{
 "Method": "EnterPIN"
"Message": "Please Enter PIN:"
"Time": "1100 hours"
 "Date": 13-5-2012
 }

**3.6.2.4. Response message:**
{
 "Method": "EnterPIN"
 "Result": "PIN",
"Time": 1100 hours

"Date": 13-5-2012"
"Error": "null"
}

### 3.6.2.5. Verification message:

```
{
 "Method": "VerifyPIN",
"Message": "Verifying PIN entry",
"Time": "1100 hours",
 "Date": "13-5-2012"
}
```

### 3.6.2.6. Response message#1:

```
{
 "Method": "VerifyPIN"
 "Result": "Verified",
"Time": "1100 hours",
 "Date": "13-5-2012"
"Error": "null"
}
```

### 3.6.2.7. Response message#2:

```
{
 "Method": "VerifyPIN"
 "Result": "NotVerified",
"Time": "1100 hours",
 "Date": "13-5-2012"
"Error": "Verification Unsuccessful"}
```

### 3.6.2.8. Transmission message:

{
 "Method": "Transmit"
 "To Account": "User No.1"
 "From Account": "User No.2"
 "Amount": "Rs. 100"
 "Time": "1100 hours"
 "Date": 13-5-2012
 }

### 3.6.2.9. Response message#1:

{
 "Method": "Transmit"
 "Result": "Transmission Successful",
"To Account": "User No.1",
 "From Account": "User No.2",
 "Amount": "Rs. 100",
 "Time": "1100 hours",
 "Date": "13-5-2012"
"Error": "null"
 }

### 3.6.2.10. Response message#2:

{
 "Method": "Transmit"
 "Result": "NotTransmitted",
"To Account": "User No.1",
"From Account": "User No.2",
 "Amount": "Rs. 100",
 "Time": "1100 hours",
 "Date": "13-5-2012"

"Error": "Account Not Found"
}

### 3.6.2.11 Response message#3:

{
 "Method": "Transmit"
 "Result": "NotTransmitted",
"To Account": "User No.1",
 "From Account": "User No.2",
 "Amount": "Rs. 100",
 "Time": "1100 hours",
 "Date": "13-5-2012",
"Error": "Account Has Not Enough Cash"
}

### 3.6.3. Cash Deposit:

### 3.6.3.1. Enter amount message:
{
 "method": "EnterAmount"
 "To Account": "User No.1"
 "From Account": " currency detector"
 "Time": "1100 hours"
 "Date": "13-5-2012"
}

### 3.6.3.2. Response:
{
"Method": "EnterAmount",
 "Result" : "Okay",
 "To Account": User No.1

"From Account": "currency detector",
"Error": "null",
"Time": 1100 hours
"Date": 13-5-2012}

### 3.6.3.3. Enter PIN message:
{
"Method": "EnterPIN"
"Message": "Please Enter PIN:"
"Time": "1100 hours"
"Date": 13-5-2012
}

### 3.6.3.4. Response message:
{
"Method": "EnterPIN"
"Result": "PIN",
"Time": 1100 hours
"Date": 13-5-2012"
"Error": "null"
}

### 3.6.3.5. Verification message:
{
"Method": "VerifyPIN",
"Message": "Verifying PIN entry",
"Time": "1100 hours",
"Date": "13-5-2012"
}

### 3.6.3.6. Response Message#1:

```
{
 "Method": "VerifyPIN"
 "Result": "Verified",
"Time": "1100 hours",
 "Date": "13-5-2012"
"Error": "null"
 }
```

### 2) Response Message#2:

```
{
 "Method": "VerifyPIN"
 "Result": "NotVerified",
"Time": "1100 hours",
 "Date": "13-5-2012"
"Error": "Verification Unsuccessful" }
```

### Transmission Message:

```
{
 "Method": "Transmit"
 "To Account": "User No.1"
 "From Account": "User No.2"
 "Amount": "Rs. 100"
 "Time": "1100 hours"
 "Date": 13-5-2012
 }
```

### 1) Response Message:

```
{
```

```
 "Method": "Transmit"
 "Result": "OK",
"To Account": "User No.1",
 "From Account": "currency detector",
 "Amount": "Rs. 100",
 "Time": "1100 hours",
 "Date": "13-5-2012"
"Error": "null"
 }
```

**2) Response Message:**

```
{
 "Method": "Transmit"
 "Result": "NotTransmitted",
"To Account": "User No.1",
 "From Account": "currency detector",
 "Amount": "Rs. 100",
 "Time": "1100 hours",
 "Date": "13-5-2012"
"Error": "Account Not Found"
 }
```

**Cash Withdrawal:**

**Enter Amount Message**:

```
{
 "method": "EnterAmount"
 "To Account": "Currency Detector"
 "From Account": "User No.1"
 "Time": "1100 hours"
 "Date": "13-5-2012"}
```

**Response:**

{
"Method": "EnterAmount",
"Result" : "Okay",
"To Account": "currency detector",
"From Account": "User No.1",
"Error": "null",
"Time": 1100 hours
"Date": 13-5-2012}

**Enter PIN Message:**

{
"Method": "EnterPIN"
"Message": "Please Enter PIN:"
"Time": "1100 hours"
"Date": 13-5-2012
}

**Response Message:**

{
"Method": "EnterPIN"
"Result": "PIN",
"Time": 1100 hours
"Date": 13-5-2012"
"Error": "null"
}

**Verification Message:**

{

```
"Method": "VerifyPIN",
"Message": "Verifying PIN entry",
"Time": "1100 hours",
"Date": "13-5-2012"
}
```

**1)Response Message#1:**
```
{
"Method": "VerifyPIN"
"Result": "Verified",
"Time": "1100 hours",
"Date": "13-5-2012"
"Error": "null"
}
```

**2)Response Message#2:**
```
{
"Method": "VerifyPIN"
"Result": "NotVerified",
"Time": "1100 hours",
"Date": "13-5-2012"
"Error": "Verification Unsuccessful" }
```

**Transmission Message:**
```
{
"Method": "Transmit"
"To Account": "currency detector"
"From Account": "User No.1"
"Amount": "Rs. 100"
"Time": "1100 hours"
```

"Date": 13-5-2012
}

**1)Response Message:**

{
 "Method": "Transmit"
 "Result": "OK",
"To Account": "currency detector",
 "From Account": "User No.1",
 "Amount": "Rs. 100",
 "Time": "1100 hours",
 "Date": "13-5-2012"
"Error": "null"
}

**2)Response Message:**

{
 "Method": "Transmit"
 "Result": "NotTransmitted",
"To Account": "currency detector",
 "From Account": "User No.1",
 "Amount": "Rs. 100",
 "Time": "1100 hours",
 "Date": "13-5-2012",
"Error": "Account Has Not Enough Cash"
}

## Hardware:

The hardware part of this system basically acts as the link between user and the software system module being employed for processing and management. It consists of different hardware modules interacting together to provide the user with various functionalities.

## Components:

The components used in the hardware module are as follows:

### 1. Arduino Mega Board:

The Arduino Mega is a microcontroller board based on the ATmega1280. It has
- 54 digital input/output pins (which can be used as PWM outputs)
- 16 analog inputs
- 4 UARTs (hardware serial ports)
- 16 MHz crystal oscillator
- USB connection
- power jack
- ICSP header
- reset button.

It essentially contains everything needed to support the microcontroller enabling the user to simply connect it to a computer with a USB cable or power it with a AC-to-DC adapter or battery to get started.

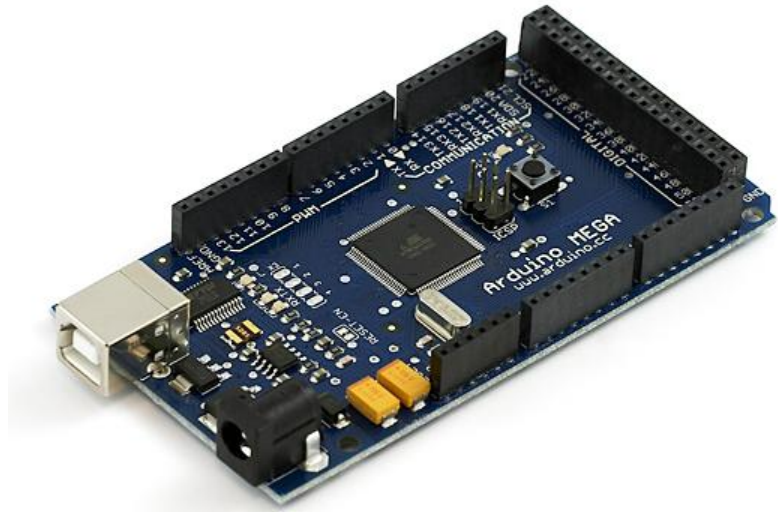

**Figure 3.12: Arduino Mega Board**

## 2. Power Supply:

The power supply Circuitry is used to provide a constant variable electric supply to all the components so as to ensure proper working and performance of all components. It consists of an old PC power supply along with voltage divider circuit to provide DC supply at 5V, 9V and 12V.

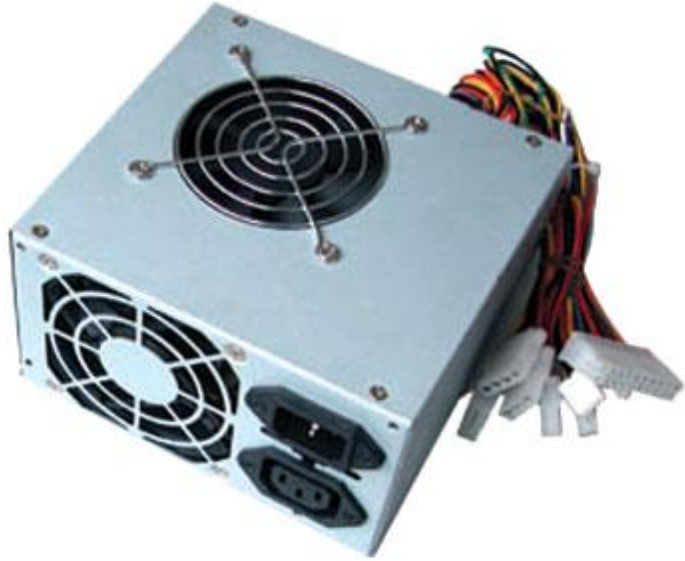

**Figure 3.13: Power Supply**

## 3. Cash Detector P77:

The Cash Detector P77 is a product by ICT Taiwan used in ATMs and for Cash Detection and acceptance worldwide. Its uses UV spectrum to analyse and decode currency notes and then store them in stack formation at the rear in an extension product.

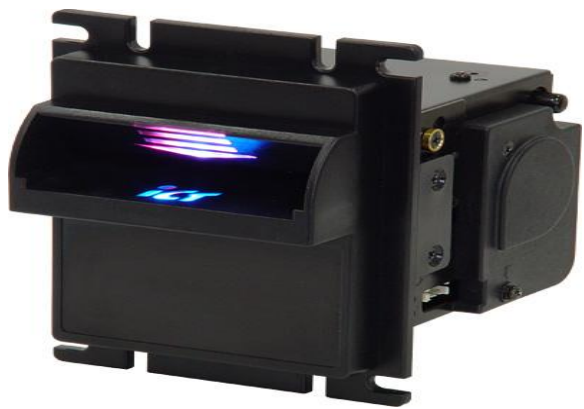

**Figure 3.14: Cash Detector**

## 4. LCD

LCD is used to make the system more interactive and offer options to the user about the system. LCDs of configuration 16*2 have been used for this purpose.

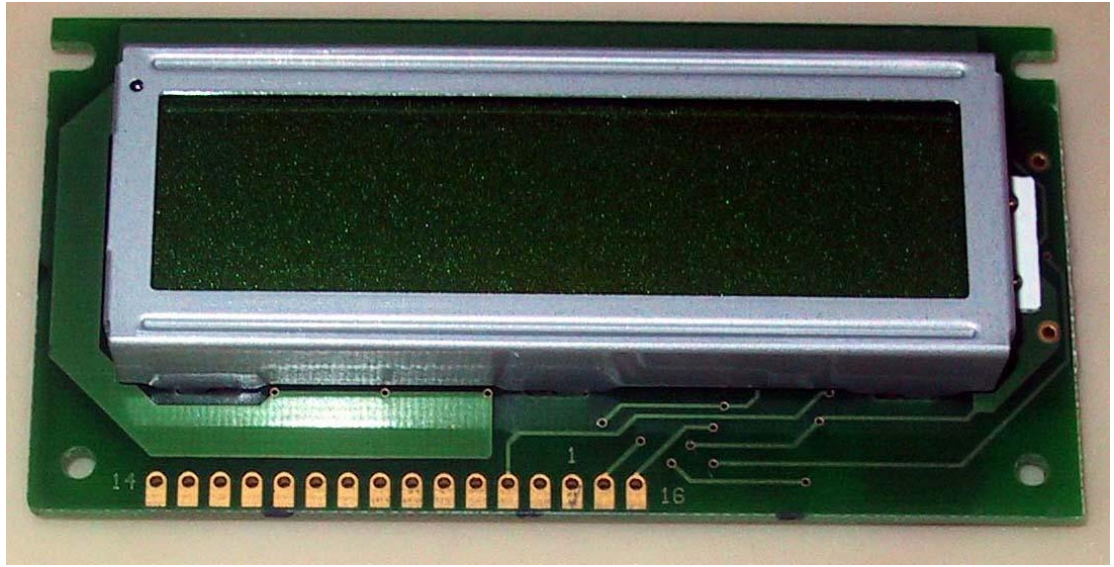

**Figure 3.15: LCD**

## 5. Smart card reader/writer:

Smart Card reader/writer is employed to give the user facility of interacting using their smart card as a debit card with the entire system and accessing their accounts and performing transactions. The smart card reader used Precise Biometrics 250 MC device.

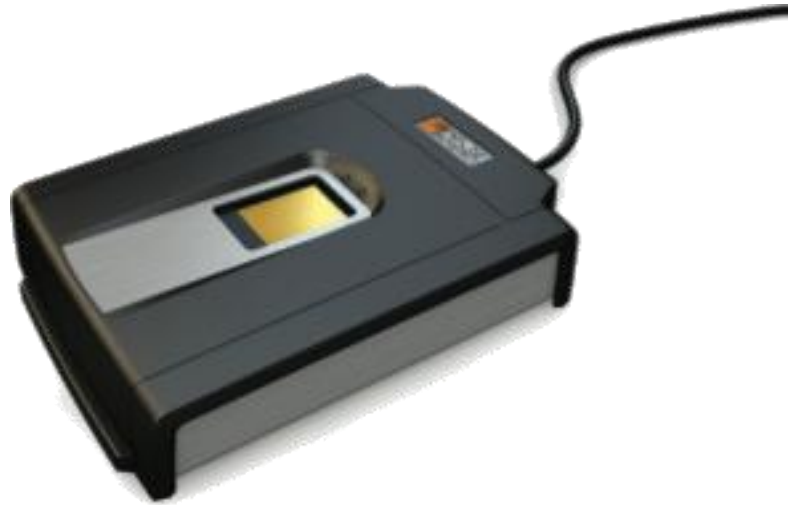

**Figure 3.16: Smart Card Reader/Writer**

## 6. MAX232 Logic Level Converter Circuit:

The information coming through the cash detector is in CMOS logic which has to be converted to TTL logic for transmission and decoding in Arduino board for which MAX232 logic level conversion Circuit has been used.

## 7. Arduino Shields:

Arduinos are capable of interacting with many devices but for some they require add-on boards known as shields. For Lan connections and USB connections Ethernet Shield and USB shield were used.

# Chapter 4

# Results

## 4.1 Architecture:

The software has two types of architectures:

- Software.
- Hardware.

### 4.1.1. Software Module:

This module consists of a client GUI which takes input data from client, sends the data to the web service at the user end. The web service at the server end takes the data, processes it in the database at back end.

### 4.1.2. GUI Interface:
#### 4.1.2.1.    Authentication:

The interface first initiates an authentication process at the terminal.

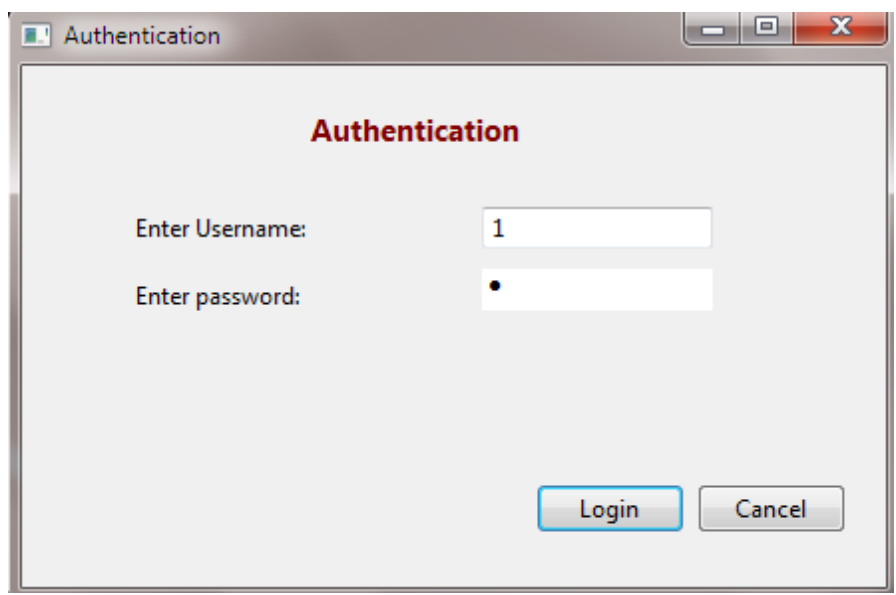

**Figure 4.1 Authentication**

### 4.1.2.2. Main menu:

The main menu displayed at all times has the following components.

- Pay Over The Counter.
- Account To Account Transfer.
- Cash Withdrawal..
- Cash Deposit.
- Add Customer
- Verify Account.
- Verify PIN.
- Cancel Transaction.

Any of these, when chosen take, the user to the relevant window of the option that the user chooses.

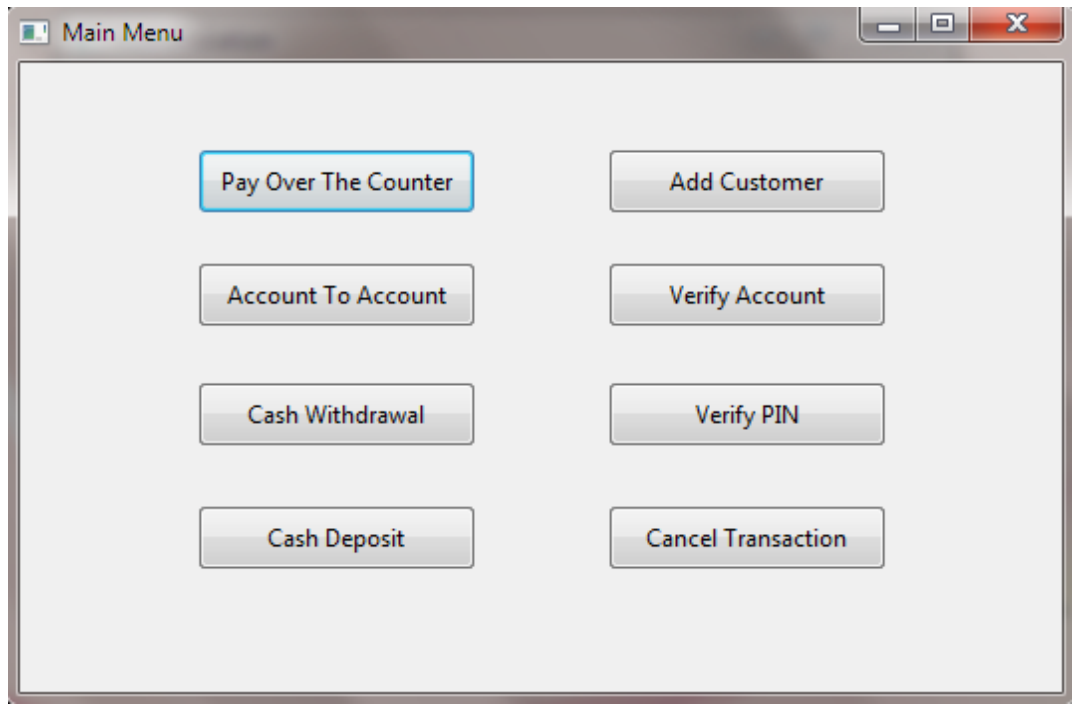

**Figure 4.2: Main Menu**

### 4.1.2.3. Pay over the counter

When the button is selected, the widget takes the user to the "Pay Over The Counter" GUI. The user is asked to insert his PIN, the account number of the transferee and the amount that the user wants to be transferred. In such an option, the customer will be paying some amount over the PoS terminal. The merchant will use his PIN to send the amount equivalent to what the customer handed to another account.

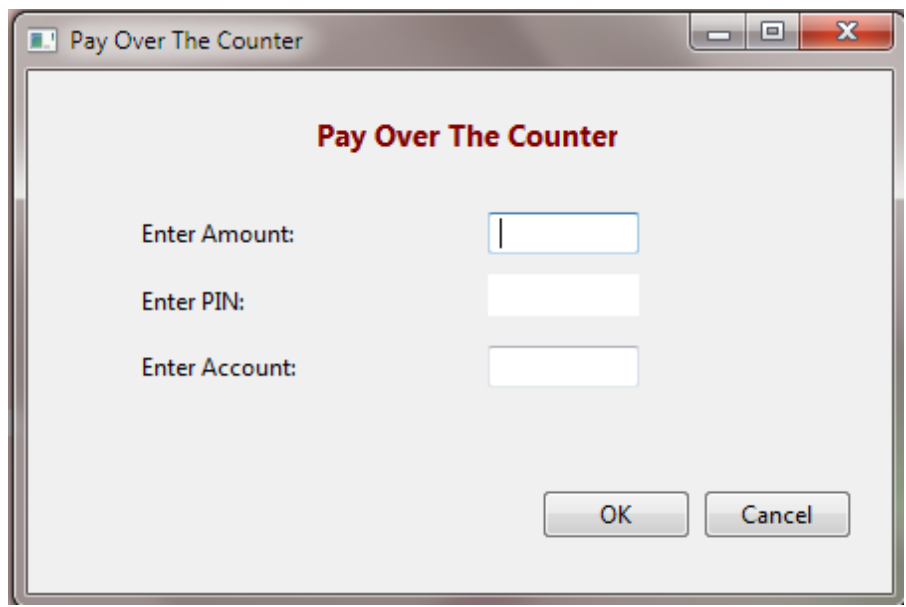

**Figure 4.3: Pay Over The Counter GUI**

### 4.1.2.4. Account to account:

When the button is selected, the widget takes the user to the "Account To Account" GUI. The user is asked to insert his PIN, the account number of the transferee and the amount that the user wants to be transferred. In such an option, the customer will be paying the merchant over the PoS terminal. The customer will use his PIN to send the amount to the merchant's account.

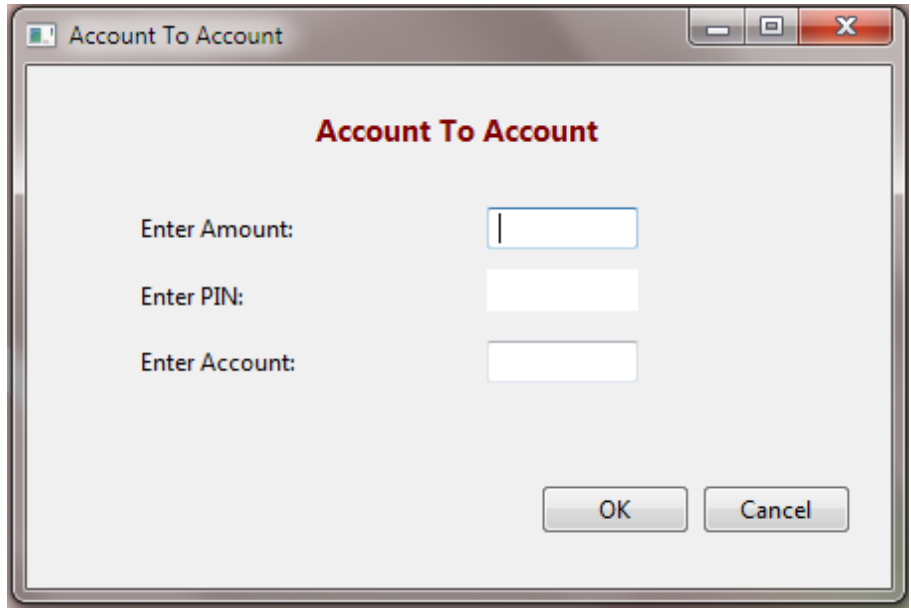

**Figure 4.4: Account To Account GUI**

### 4.1.2.5. Cash withdrawal:

When the button is selected, the widget takes the user to the "Cash Withdrawal" GUI. The merchant will have the user enter his PIN and transfer the said amount into the merchant's account. The merchant then hands over the cash to the user.

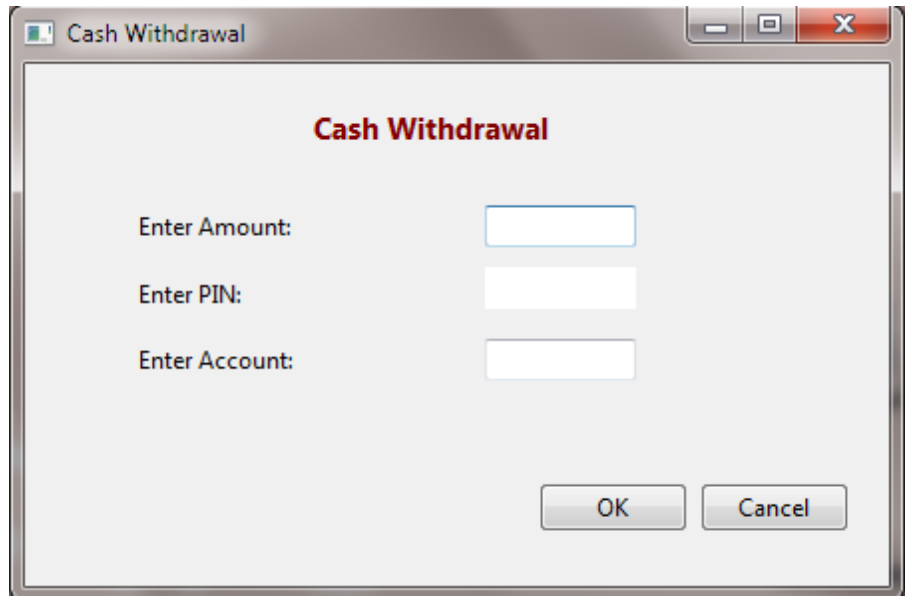

**Figure 4.5: Cash Withdrawal GUI**

### 4.1.2.6. Cash deposit:

When the button is selected, the widget takes the user to the "Cash Deposit" GUI. The merchant will insert his PIN and transfer the amount equivalent to what the customer hands the merchant. The customer's account will be entered.

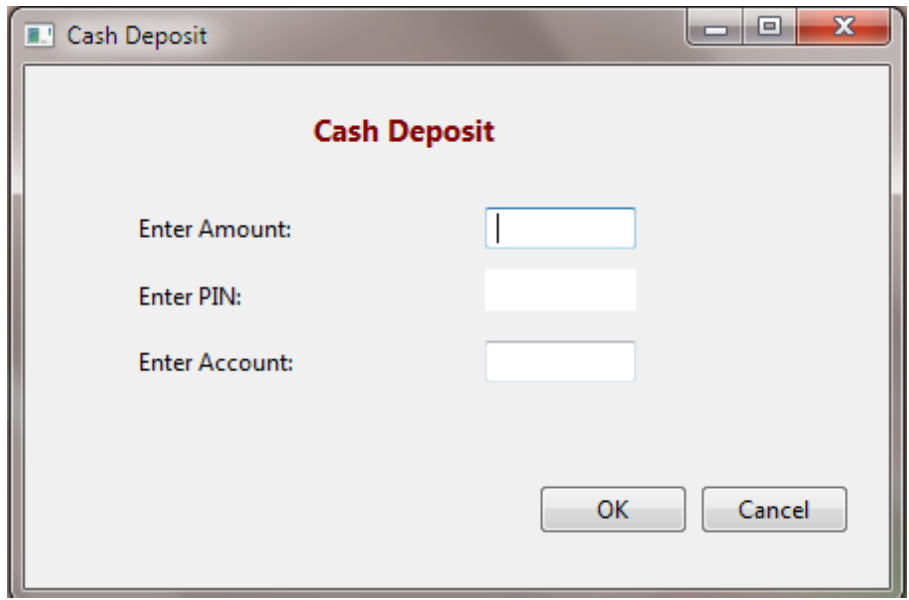

**Figure 4.6: Cash Deposit**

### 4.1.2.7. Add Customer

A customer maybe added to the backend database via web service.

**Figure 4.7: Add Customer**

### 4.1.2.8. Verify account:

The user may enter his account number to check if it exists in the system before he may proceed to make a transaction.

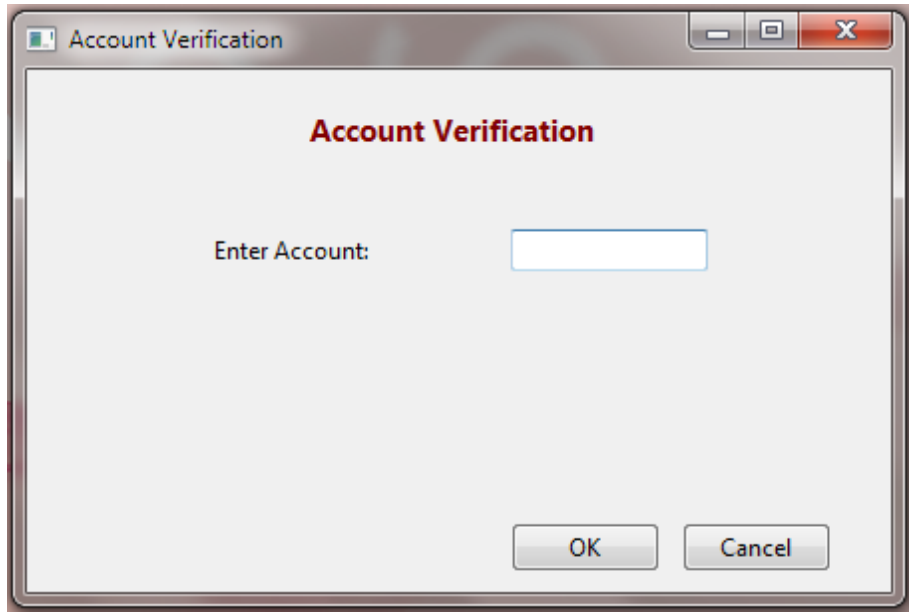

**Figure 4.8: Account Verification**

### 4.1.2.9. PIN verification:

The user may enter his PIN to check if his details exist in the system.

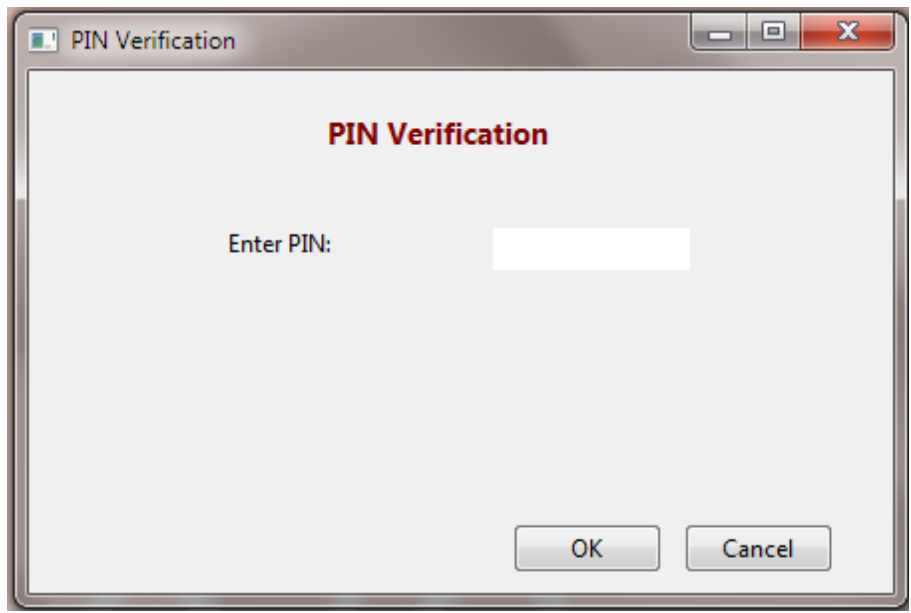

**Figure 4.9: PIN Verification**

### 4.1.2.10. Cancel transaction:

The user can cancel a transaction by entering the amount he transferred previously to another account. The user will enter his PIN and the account number of the person, to whom the amount was sent. Cash will be recovered in the user's account.

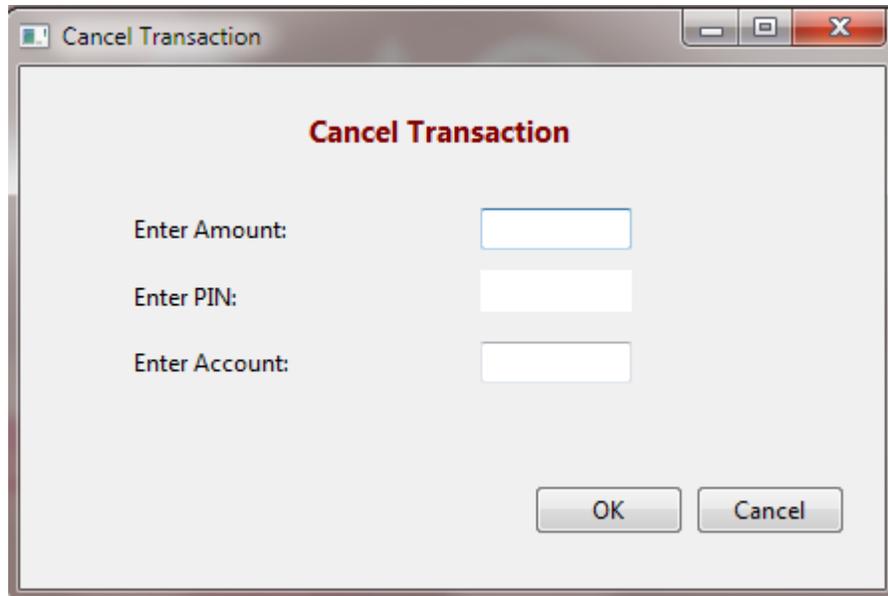

**Figure 4.10: Cancel Transaction**

### 4.1.3. How The System Is Working:

A web service is running. The web service processes the user request. A transaction function is invoked whenever a user selects any of the payment options. With sufficient input information from the user, the web service aids in processing the data at the back end. The mathematical manipulations are done in the database at the back end. There are functions for at the back running behind every button for processing user's request. Please refer to the Appendix A for the code.

### 4.1.4. Web Service:

The web service is the building and the binding block of the software module. It creates a communication channel between the client and server. This channel is secure. The web services implemented in the project are run on Apache Tomcat v6. For the web service code of the project, please refer to the Appendix A.

### 4.1.5. Database:

The database has been implemented in MySQL Workbench 5.2 CE. The following Entity-Relationship Diagram aids in connections between different components of the project.

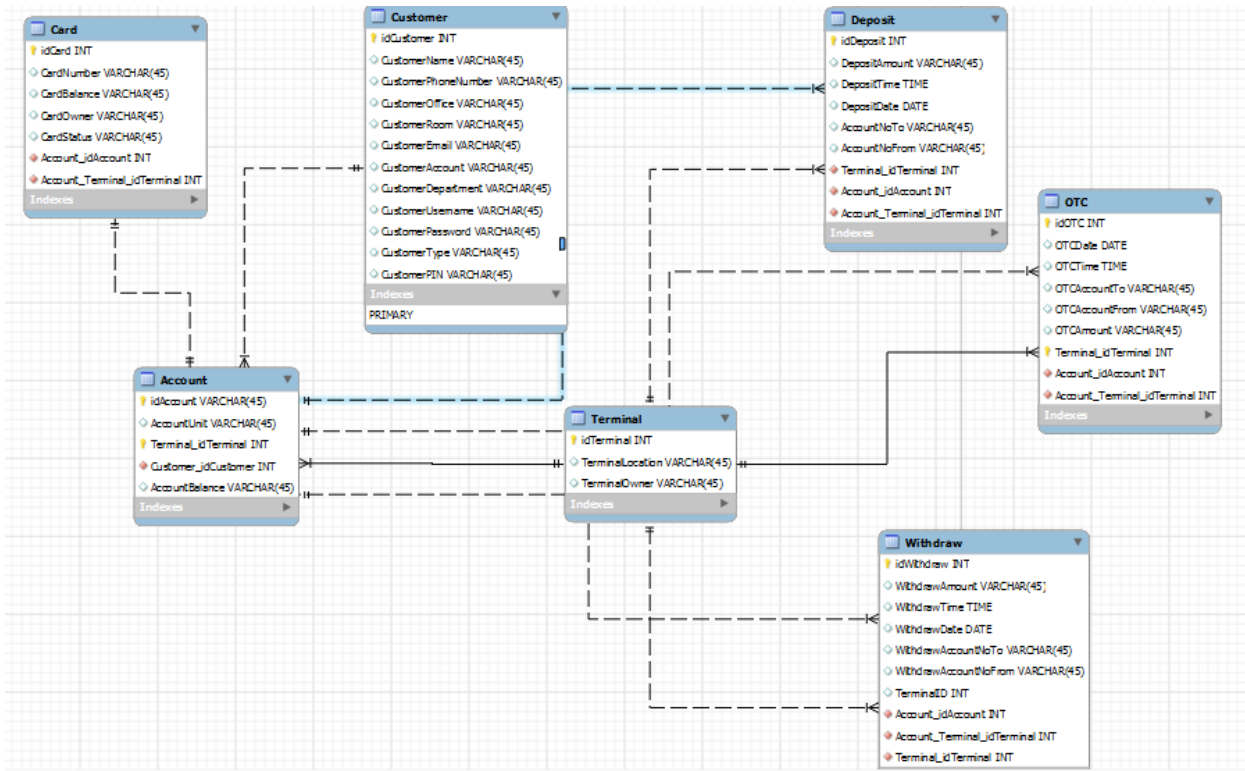

**Figure 4.11: Backend Database Tables in MySQL Workbench 5.2 CE**

### 4.1.6. Hardware Module:

An embedded device is developed which has certain modules synchronized with each other to provide functionality. The user will insert the smart card into the embedded system via card reader. The system will get in touch with the backend server for authenitcation and authorization wiith the help of WLAN connection. Once the connection is made, mathematical manipulations of cost deduction are done on the card as well as in synchronum, at the backend servers.

### 4.1.7. Hardware Methodology:
The hardware module's methodology can be divided into:

1. **Smart card entry and verification:**

    In this cycle the smart card is entered and user verifies his identity using PIN

2. **Option selection:**

    Options are made available on the LCD and user selects it from them to make transactions or avail any other option.

3. **Cash Detector cycle:**

    In this cycle cash is entered into the cash detector and is loaded on to the smart card.

4. **Synchronization:**

    All the information that is being exchanged is also being updated on the back end server through the network connection so that the information in the smartcard and the server is synchronised.

# Chapter 5

# Discussion And Future Work

The project implements the transaction functions so far. Further features can be added to the project. The system has been designed for small-scale transactions for a small area. The system can be implemented on a large scale enterprise level by adding further security features. Multiple points of the system can be constructed with all the points connected together with a larger server.

64 | P a g e## REFERENCES:

64 | P a g e

## REFERENCES:

[1] "What are chip-enabled EMV payment cards?"

http://www.smartcardalliance.org/pages/slideshows-20120409?template=slides

[2] "Smart Card Technology and the National Cybersecurity Strategy"

http://www.smartcardalliance.org/pages/slideshows-20101111?template=slides

[3] "Benefits of Smart Cards versus Magnetic Stripe Cards for Healthcare Applications"

http://www.smartcardalliance.org/resources/pdf/smart_cards_vs_mag_stripe_cards_122111.pdf

[4] "Smart Card Tutorial - Part 1 First Published in September 1992"

http://www.smartcard.co.uk/tutorials/sct-itsc.pdf

# APPENDIX A

## Web Service Code

**Imported Classes:**

import java.io.IOException;

import java.sql.Connection;

import java.sql.DriverManager;

import java.sql.ResultSet;

import java.sql.ResultSetMetaData;

import java.sql.SQLException;

import java.sql.Statement;

import java.util.Calendar;

import java.util.Vector;

import com.cedarsoftware.util.io.JsonReader;

import com.cedarsoftware.util.io.JsonWriter;

**Web Service Class "JunoS":**

public class JunoS{

    static Connection conn;

    static Statement state;

    static ResultSet rs;

```java
    public void connection() {

        try {

            Class.forName("com.mysql.jdbc.Driver");

        } catch (ClassNotFoundException ex) {

            System.out.println("class not found");

        }     try {

    conn = (Connection) DriverManager.getConnection("jdbc:mysql://localhost/mydb?" + "user=root&password=rumi");

        } catch (SQLException ex) {

            System.out.println("sql exception");

        }

        try {

            state = (Statement) conn.createStatement();

            System.out.println("Connection Successful");

        } catch (SQLException ex) {

            System.out.println("Connection Unsuccessful");

        }   }
   int insertData(String q) {

        try {

    state.executeUpdate(q);

            System.out.println("Inserted");
```

```java
            int success = 1;

            return success;       } catch (SQLException ex) {

            System.out.println("Cannot Insert");

            int error = 2;

            return error; } }

    public void searchData(String query) throws IOException {

        //String query = "select * from user ";

        Vector<String> in=new Vector<String>(50,50);

        Vector<Object> out=new Vector<Object>(20,20);

try {

            rs = state.executeQuery(query);

            ResultSetMetaData rsmd = rs.getMetaData();

            int cols = rsmd.getColumnCount();

        System.out.println("Column Count is " + cols);

           int j=0;

            while (rs.next())

            {

                for(int i=1;i<=cols;i++)

                {

                    String s=rs.getString(i);

                    in.add(s) ;
```

```java
                System.out.println(in.get(j));
                j++;
            }
            out.add(in);
            System.out.println(out.get(0));
        }
//      int success = 1;
//      return success;
        } catch (SQLException ex) {
            Calendar cal = Calendar.getInstance();
            String b = "\n" + "MessageType: Error" + "\n" +
            "Result: Not Found" + "\n" +
                cal.getTime() + "\n";
                String json3 = JsonWriter.objectToJson(b); //Balance Not Sufficient
//              System.out.println(json3);
                Object json4 = JsonReader.jsonToJava(json3);
                System.out.println(json4);
//              JunoS d = new JunoS();
//              d.start();
//              int error = 2;
//      return error;  } }
```

```java
public void insertdata(String args[])
    {   JunoS d=new JunoS();
            d.insertData("insert into terminal(idCard, CardNumber, CardBalance, CardOwner, CardStatus) values (\"1\",\"1\",\"120\",\"Rum\",\"Active\");");}

    void updatedata(String query)
    {   java.sql.Statement stmt = null;
            try {
stmt = conn.createStatement();
            } catch (SQLException e1) {
                // TODO Auto-generated catch block
                e1.printStackTrace();}
try {
                stmt.executeUpdate(query);
            } catch (SQLException e) {
                // TODO Auto-generated catch block
                e.printStackTrace();}
   System.out.println("Updated");   }

    public int addcustomer(String a, String b, String c, String d, String e, String f, String g, String h, String i, String j, String k)
    {   int value =0;
        value = insertData("insert into customer(idCustomer, CustomerName, CustomerPhoneNumber, CustomerOffice, CustomerRoom, CustomerEmail, CustomerAccount,
```

```java
CustomerDepartment, CustomerUsername, CustomerPassword, CustomerPIN) values
("'+a+"','"+b+"','"+c+"','"+d+"','"+e+"','"+f+"','"+g+"','"+h+"','"+i+"','"+j+"','"+k+"')");

        //int success = 1;

        //return success;

return value;

    }

  public int payoverthecounter(String amount, String pin, String account) throws IOException

    {   Calendar cal = Calendar.getInstance();

            System.out.println("You selected: 1)Pay Over The Counter");

            String b = "\n" + "Method: Enter Amount" + "\n" +

            "To Account: Merchant" + "\n" + "From Account: User" + "\n" +

            cal.getTime() + "\n";

            String json3 = JsonWriter.objectToJson(b); //Enter Amount Message

    //      System.out.println(json3);

             Object json4 = JsonReader.jsonToJava(json3);

            System.out.println(json4);

         String z = "0";

        z=amount;

        String c = ("\n" + "Amount: " + z + "\n" +

         "To Account: Merchant" + "\n" + "From Account: User" + "\n" +

         cal.getTime() + "\n");
```

```java
            String json5 = JsonWriter.objectToJson(c); //User Enters Amount
//          System.out.println(json3);
            Object json6 = JsonReader.jsonToJava(json5);
            System.out.println(json6);
String e = "\n" + "Method: Enter PIN" + "\n" +
            "To Account: Merchant" + "\n" + "From Account: User" + "\n" +
            cal.getTime() + "\n";
            String json7 = JsonWriter.objectToJson(e);
            Object json8 = JsonReader.jsonToJava(json7);
            System.out.println(json8);
            String r="0";
            r = pin;
            String f = ("\n" + "PIN: " + r + "\n" +
             "To Account: Merchant" + "\n" + "From Account: User" + "\n" +
             cal.getTime() + "\n");
            String json9 = JsonWriter.objectToJson(f); //User Enters Amount
//          System.out.println(json3);
            Object json10 = JsonReader.jsonToJava(json9);
            System.out.println(json10);
```

```java
        JunoS d=new JunoS();

        d.verifypin(r);

          d.checkbalance("SELECT * from Card WHERE CardBalance >="+z+"");

           String u="0";

//      d.searchData("Select u = idAccount from Customer WHERE PIN = "+y+"");

         String g = "\n" + "Method: Enter Recipient Account" + "\n" +

         "To Account: Merchant" + "\n" + "From Account: User" + "\n" +

         cal.getTime() + "\n";

         String json11 = JsonWriter.objectToJson(g);

         Object json12 = JsonReader.jsonToJava(json11);

         System.out.println(json12);

         String h="0";

        h=account;

       String i = ("\n" + "Recipient Account: " + h + "\n" +

         "To Account: Merchant" + "\n" + "From Account: User" + "\n" +

         cal.getTime() + "\n");

          String json13 = JsonWriter.objectToJson(i); //User Enters Amount

//      System.out.println(json3);

          Object json14 = JsonReader.jsonToJava(json13);

         System.out.println(json14);
```

```java
            d.verifypin(r);

            d.checkbalance("SELECT * from Card WHERE Account_idAccount = '"+h+"' AND CardBalance >='"+z+"'");

     int success =    d.transaction(z,r,h);

      return success;   }

     public int a2a(String amount, String pin, String account) throws IOException

    {Calendar cal = Calendar.getInstance();

            System.out.println("You selected: 2)Account To Account Transfer");

                String b = "\n" + "Method: Enter Amount" + "\n" +

                "To Account: User#1" + "\n" + "From Account: User#2" + "\n" +

                cal.getTime() + "\n";

                String json3 = JsonWriter.objectToJson(b);

   //         System.out.println(json3);

                    Object json4 = JsonReader.jsonToJava(json3);

              System.out.println(json4);

             String z="0";

             z=amount;

             String c = ("\n" + "Amount: " + z + "\n" +

              "To Account: User#1" + "\n" + "From Account: User#2" + "\n" +

              cal.getTime() + "\n");

             String json5 = JsonWriter.objectToJson(c);
```

```java
//          System.out.println(json3);

            Object json6 = JsonReader.jsonToJava(json5);

            System.out.println(json6);

            String e = "\n" + "Method: Enter PIN" + "\n" +

    "To Account: User#1" + "\n" + "From Account: User#2" + "\n" +

    cal.getTime() + "\n";

            String json7 = JsonWriter.objectToJson(e);

            Object json8 = JsonReader.jsonToJava(json7);

            System.out.println(json8);

            String y="0";

           y=pin;

String f = ("\n" + "PIN: " + y + "\n" +

            "To Account: User#1" + "\n" + "From Account: User#2" + "\n" +

            cal.getTime() + "\n");

            String json9 = JsonWriter.objectToJson(f); //User Enters Amount

//          System.out.println(json3);

            Object json10 = JsonReader.jsonToJava(json9);

            System.out.println(json10);

            String io = "\n" + "Method: Enter Account" + "\n" +

                    "To Account: User#1" + "\n" + "From Account: User#2" + "\n" +

                    cal.getTime() + "\n";
```

```java
                        String json13 = JsonWriter.objectToJson(io);

                        Object json14 = JsonReader.jsonToJava(json13);

                        System.out.println(json14);

                        String yo="0";

                       yo=account;

            String fo = ("\n" + "Account: " + yo + "\n" +

                        "To Account: User#1" + "\n" + "From Account: User#2" + "\n" +

                        cal.getTime() + "\n");

                        String json15 = JsonWriter.objectToJson(fo); //User Enters Amount

                //      System.out.println(json3);

                         Object json16 = JsonReader.jsonToJava(json15);

                        System.out.println(json16);

                        JunoS d=new JunoS();

            d.verifypin(y);

              d.checkbalance("SELECT * from Card WHERE Account_idAccount = '"+yo+"' AND CardBalance >='"+z+"'");

                int success=  d.transaction(z,y,yo);

            return success;

                 }

     public int cashwithdrawal(String s, String p, String q) throws IOException

      {
```

```java
            Calendar cal = Calendar.getInstance();

            System.out.println("You selected: 3)Cash Withdrawal");

                String b = "\n" + "Method: Enter Amount" + "\n" +

                    "To Account: Currency Detector" + "\n" + "From Account: User" + "\n" +

                    cal.getTime() + "\n";

                String json3 = JsonWriter.objectToJson(b);

//              System.out.println(json3);

                Object json4 = JsonReader.jsonToJava(json3);

                System.out.println(json4);

                String z="0";

                z = s;

                String c = ("\n" + "Amount: " + z + "\n" +

                    "To Account: Currency Detector" + "\n" + "From Account: User" + "\n" +

                    cal.getTime() + "\n");

                String json5 = JsonWriter.objectToJson(c);

//              System.out.println(json3);

                 Object json6 = JsonReader.jsonToJava(json5);

                 System.out.println(json6);

                String e = "\n" + "Method: Enter PIN" + "\n" +

            "To Account: Currency Detector" + "\n" + "From Account: User" + "\n" +
```

```java
                cal.getTime() + "\n";
        String json7 = JsonWriter.objectToJson(e);
        Object json8 = JsonReader.jsonToJava(json7);
        System.out.println(json8);
        String y="0";
    y=p;
        String f = ("\n" + "PIN: " + y + "\n" +
         "To Account: Currency Detector" + "\n" + "From Account: User" + "\n" +
        cal.getTime() + "\n");
        String json9 = JsonWriter.objectToJson(f); //User Enters Amount
//        System.out.println(json3);
         Object json10 = JsonReader.jsonToJava(json9);
        System.out.println(json10);
        String io = "\n" + "Method: Enter Account" + "\n" +
                    "To Account: User#1" + "\n" + "From Account: User#2" + "\n" +
                    cal.getTime() + "\n";
                String json13 = JsonWriter.objectToJson(io);
                Object json14 = JsonReader.jsonToJava(json13);
                System.out.println(json14);
                String yo="0";
                yo=q;
```

```java
                    String fo = ("\n" + "Account: " + yo + "\n" +
                            "To Account: User#1" + "\n" + "From Account: User#2" + "\n" +
                            cal.getTime() + "\n");
                    String json15 = JsonWriter.objectToJson(fo); //User Enters Amount
                    //      System.out.println(json3);
                    Object json16 = JsonReader.jsonToJava(json15);
                    System.out.println(json16);
                    JunoS d=new JunoS();
            d.verifypin(y);
            d.checkbalance("SELECT * from Card WHERE Account_idAccount = '"+q+"' AND CardBalance >='"+z+"'");
            int success =  d.transaction(z, y, yo);
            return success; }
        public int cashdeposit(String s, String p, String q) throws IOException
        {Calendar cal = Calendar.getInstance();
            System.out.println("You selected: 4)Cash Deposit");
                    String b = "\n" + "Method: Enter Amount" + "\n" +
                    "To Account: User" + "\n" + "From Account: Currency Detector" + "\n" +
                    cal.getTime() + "\n";

                String json3 = JsonWriter.objectToJson(b);
```

```java
//        System.out.println(json3);
        Object json4 = JsonReader.jsonToJava(json3);
        System.out.println(json4);
    String z="0";
    z = s;
    String c = ("\n" + "Amount: " + z + "\n" +
        "To Account: User" + "\n" + "From Account: Currency Detector" + "\n" +
        cal.getTime() + "\n");
        String json5 = JsonWriter.objectToJson(c);
//        System.out.println(json3);
        Object json6 = JsonReader.jsonToJava(json5);
        System.out.println(json6);
      String e = "\n" + "Method: Enter PIN" + "\n" +
 "To Account: User" + "\n" + "From Account: Currency Detector" + "\n" +
 cal.getTime() + "\n";
 String json7 = JsonWriter.objectToJson(e);
 Object json8 = JsonReader.jsonToJava(json7);
 System.out.println(json8);
 String y ="0";
y = p;
```

```java
            String f = ("\n" + "PIN: " + y + "\n" +
                "To Account: User" + "\n" + "From Account: Currency Detector" + "\n" +
                cal.getTime() + "\n");
            String json9 = JsonWriter.objectToJson(f); //User Enters Amount
//           System.out.println(json3);
            Object json10 = JsonReader.jsonToJava(json9);
            System.out.println(json10);
            String io = "\n" + "Method: Enter Account" + "\n" +
                        "To Account: User#1" + "\n" + "From Account: User#2" + "\n" +
                        cal.getTime() + "\n";
                String json13 = JsonWriter.objectToJson(io);
                Object json14 = JsonReader.jsonToJava(json13);
                System.out.println(json14);
                String yo="0";
                yo=q;
            String fo = ("\n" + "Account: " + yo + "\n" +
                    "To Account: User#1" + "\n" + "From Account: User#2" + "\n" +
                    cal.getTime() + "\n");
             String json15 = JsonWriter.objectToJson(fo); //User Enters Amount
//              System.out.println(json3);
                  Object json16 = JsonReader.jsonToJava(json15);
```

```java
                    System.out.println(json16);

                    JunoS d=new JunoS();

            int success1 = d.verifypin(y);

//          return success1;

                d.checkbalance("SELECT * from Card WHERE  Account_idAccount = '"+q+"' AND CardBalance >='"+z+"'");

            int success =   d.transaction(z,y,yo);

            return success;}

        public int verifypin(String z) throws IOException

        {JunoS d=new JunoS();

    d.checkpin("select * from Customer where CustomerPIN = '"+z+"'");

     int success = 1;

      return success;}

        public  void verifyaccount(String account) throws IOException

        {JunoS d=new JunoS();

    d.checkaccount("select * from Account where idAccount = '"+account+"'");}

        public int transaction(String amount, String pin, String account) throws IOException

        {JunoS d = new JunoS();

                d.updatedata("update Card set CardBalance = CardBalance+ '"+amount+"' where Account_idAccount = '"+account+"'");

                d.updatedata("update Account set AccountUnit = AccountUnit+ '"+amount+"' where idAccount = '"+account+"'");
```

```java
            d.updatedata("update customer, account SET customer.CustomerBalance = account.AccountUnit where idCustomer = (Select Customer_idCustomer from account where idAccount = '"+account+"')");

        d.updatedata("update Customer set CustomerBalance = CustomerBalance-'"+amount+"' where CustomerPIN = '"+pin+"'");

            d.updatedata("UPDATE account, customer SET account.AccountUnit = CustomerBalance where Customer_idCustomer = '"+pin+"'");

        //      d.updatedata("UPDATE card, account SET card.CardBalance = account.AccountUnit where Account_idAccount = ;

            d.updatedata("update card, account set card.CardBalance = account.AccountUnit where Account_idAccount = (Select idAccount  from account where Customer_idCustomer = '"+pin+"')");

            int success = 1;

            return success;}
public int canceltransaction(String amount, String pin, String account) throws IOException

    {JunoS d = new JunoS();

            d.updatedata("update Card set CardBalance = CardBalance-'"+amount+"' where Account_idAccount = '"+account+"'");

            d.updatedata("update Account set AccountUnit = AccountUnit- '"+amount+"' where idAccount = '"+account+"'");

            d.updatedata("update customer, account SET customer.CustomerBalance = account.AccountUnit where idCustomer = (Select Customer_idCustomer from account where idAccount = '"+account+"')");
```

```
            d.updatedata("update Customer set CustomerBalance = CustomerBalance+'"+amount+"' where CustomerPIN = '"+pin+"'");

            d.updatedata("UPDATE account, customer SET account.AccountUnit = CustomerBalance where Customer_idCustomer = '"+pin+"'");

    //      d.updatedata("UPDATE card, account SET card.CardBalance = account.AccountUnit where Account_idAccount = ;

    d.updatedata("update card, account set card.CardBalance = account.AccountUnit where Account_idAccount = (Select idAccount  from account where Customer_idCustomer = '"+pin+"')");

int success = 1;

            return success;}

public int checkbalance(String query) throws IOException

    {    //String query = "select * from user ";

    Vector<String> in=new Vector<String>(50,50);

    Vector<Object> out=new Vector<Object>(20,20);

    try {

            rs = state.executeQuery(query);

            ResultSetMetaData rsmd = rs.getMetaData();

            int cols = rsmd.getColumnCount();

            System.out.println("Column Count is " + cols);

            int j=0;

            while (rs.next())

            {   for(int i=1;i<=cols;i++)
```

```java
            {
                String s=rs.getString(i);
                in.add(s) ;
                System.out.println(in.get(j));
                j++;
            }
            out.add(in);
            System.out.println(out.get(0));
        }
            int success = 1;
            return success;
        } catch (SQLException ex) {
            Calendar cal = Calendar.getInstance();
            String b = "\n" + "MessageType: Error" + "\n" +
            "Result: Balance Not Sufficient" + "\n" +
                    cal.getTime() + "\n";
    String json3 = JsonWriter.objectToJson(b); //Balance Not Sufficient
      //       System.out.println(json3);
                Object json4 = JsonReader.jsonToJava(json3);
                System.out.println(json4);
      //        JunoS d = new JunoS();
```

```java
                    int error = 2;
                    return error;} }
public int checkpin(String query) throws IOException
    {  //String query = "select * from user ";
     Vector<String> in=new Vector<String>(50,50);
     Vector<Object> out=new Vector<Object>(20,20);
     try {
            rs = state.executeQuery(query);
            ResultSetMetaData rsmd = rs.getMetaData();
            int cols = rsmd.getColumnCount();
            System.out.println("Column Count is " + cols);
            int j=0;
            while (rs.next())
        { for(int i=1;i<=cols;i++)
            {
             String s=rs.getString(i);
               in.add(s) ;
                System.out.println(in.get(j));
              j++;
            }
            out.add(in);
```

```java
            System.out.println(out.get(0));

        }

           int success = 1;

           return success;

        } catch (SQLException ex) {

            Calendar cal = Calendar.getInstance();

            String b = "\n" + "MessageType: Error" + "\n" +

            "Result: PIN not verified" + "\n" +

                 cal.getTime() + "\n";

         String json3 = JsonWriter.objectToJson(b); //Balance Not Sufficient

         //        System.out.println(json3);

                  Object json4 = JsonReader.jsonToJava(json3);

                  System.out.println(json4);

         //        JunoS d = new JunoS();

                  int error = 2;

                  return error;  }   }
    public int checkaccount(String query) throws IOException

    {

//String query = "select * from user ";

    Vector<String> in=new Vector<String>(50,50);

    Vector<Object> out=new Vector<Object>(20,20);
```

```java
try {

    rs = state.executeQuery(query);

    ResultSetMetaData rsmd = rs.getMetaData();

    int cols = rsmd.getColumnCount();

    System.out.println("Column Count is " + cols);

    int j=0;

    while (rs.next())

    {

    for(int i=1;i<=cols;i++)

    {

     String s=rs.getString(i);

       in.add(s) ;

         System.out.println(in.get(j));

        j++;

    }            out.add(in);

    System.out.println(out.get(0));

  }

    int success = 1;

    return success;

  } catch (SQLException ex) {

    Calendar cal = Calendar.getInstance();
```

```java
        String b = "\n" + "MessageType: Error" + "\n" +
            "Result: Account Not Available" + "\n" +
                cal.getTime() + "\n";
        String json3 = JsonWriter.objectToJson(b); //Balance Not Sufficient
//        System.out.println(json3);
         Object json4 = JsonReader.jsonToJava(json3);
        System.out.println(json4);
//        JunoS d = new JunoS();
        int error = 2;
        return error;}}
```

## Authentication

```java
package localhost.axis.JunoS_jws;

import java.rmi.RemoteException;

import javax.xml.rpc.ServiceException;

import org.eclipse.swt.widgets.Display;

import org.eclipse.swt.widgets.Shell;

import org.eclipse.swt.widgets.Text;

import org.eclipse.swt.SWT;

import org.eclipse.swt.widgets.Label;

import org.eclipse.swt.widgets.Button;
```

```java
import org.eclipse.swt.events.SelectionAdapter;

import org.eclipse.swt.events.SelectionEvent;

import org.eclipse.swt.graphics.Color;

import org.eclipse.swt.graphics.Device;

import org.eclipse.swt.widgets.Canvas;

import org.eclipse.wb.swt.SWTResourceManager;

public class Authentication {

    protected Shell shell;

    private Text text;

    private Text text_1;

public String x;

    public String y;

    /**
     * Launch the application.
     * @param args
     */
    public static void main(String[] args) {

        try {

            Authentication window = new Authentication();

            window.open();

        } catch (Exception e) {
```

```
                e.printStackTrace();}}/**
 * Open the window.
 */
public void open() {
        Display display = Display.getDefault();
        createContents();
        shell.open();
        shell.layout();
        while (!shell.isDisposed()) {
                if (!display.readAndDispatch()) {
                        display.sleep();}}}/**
 * Create contents of the window.
 */
protected void createContents() {
        shell = new Shell();
        shell.setSize(450, 300);
        shell.setText("Authentication");
        text = new Text(shell, SWT.BORDER);
        text.setBounds(231, 71, 116, 21);
        text_1 = new Text(shell, SWT.PASSWORD);
        text_1.setBounds(231, 102, 116, 21);
```

```java
        Button btnOk = new Button(shell, SWT.NONE);
        btnOk.addSelectionListener(new SelectionAdapter() {
            @Override
            public void widgetSelected(SelectionEvent arg0) {
                x = text.getText();
                y = text.getText();
    if (x.equals("1") && y.equals("1"))
    {
        Main window = new Main();
        window.open();
    }
    else
    {
        ErrorMessage window = new ErrorMessage();
        window.open();
}}});
        btnOk.setBounds(258, 210, 75, 25);
        btnOk.setText("Login");
        Label lblEnterAmount = new Label(shell, SWT.NONE);
        lblEnterAmount.setBounds(57, 74, 86, 15);
        lblEnterAmount.setText("Enter Username:");
```

```java
        Label lblEnterPin = new Label(shell, SWT.NONE);

        lblEnterPin.setBounds(57, 108, 86, 15);

        lblEnterPin.setText("Enter password:");

        Button btnCancel = new Button(shell, SWT.NONE);

        btnCancel.addSelectionListener(new SelectionAdapter() {

            @Override

            public void widgetSelected(SelectionEvent arg0) {

                shell.close();}});

        btnCancel.setBounds(339, 210, 75, 25);

        btnCancel.setText("Cancel");

        Label lblCancelTransaction = new Label(shell, SWT.NONE);

        lblCancelTransaction.setFont(SWTResourceManager.getFont("Segoe UI", 11, SWT.BOLD));

        lblCancelTransaction.setForeground(SWTResourceManager.getColor(SWT.COLOR_DARK_RED));

        lblCancelTransaction.setBounds(145, 22, 151, 21);

        lblCancelTransaction.setText("Authentication");}}
```

## Main

```java
package localhost.axis.JunoS_jws;

import org.eclipse.swt.widgets.Display;

import org.eclipse.swt.widgets.Shell;

import org.eclipse.swt.widgets.Button;
```

```java
import org.eclipse.swt.SWT;

import org.eclipse.swt.events.SelectionAdapter;

import org.eclipse.swt.events.SelectionEvent;

import localhost.axis.JunoS_jws.PayOverTheCounter;

import localhost.axis.JunoS_jws.AccountToAccount;

import localhost.axis.JunoS_jws.CashWithdrawal;

public class Main {

protected Shell shell;

    /**
     * Launch the application.
     * @param args
     */
    public static void main(String[] args) {

        try {

            Main window = new Main();

            window.open();

        } catch (Exception e) {

            e.printStackTrace();}}/**
     * Open the window.
     */
    public void open() {
```

```java
        Display display = Display.getDefault();

        createContents();

        shell.open();

        shell.layout();

        while (!shell.isDisposed()) {

            if (!display.readAndDispatch()) {

                display.sleep();}}}    /**
 * Create contents of the window.
 */
protected void createContents() {

    shell = new Shell();

    shell.setSize(539, 354);

    shell.setText("Main Menu");

    Button btnPayOverCounter = new Button(shell, SWT.NONE);

    btnPayOverCounter.addSelectionListener(new SelectionAdapter() {

        @Override

        public void widgetSelected(SelectionEvent arg0) {

            try {

                PayOverTheCounter window = new PayOverTheCounter();

                window.open();

            } catch (Exception e) {
```

```java
                e.printStackTrace();
        }}});
btnPayOverCounter.setBounds(89, 43, 140, 33);
btnPayOverCounter.setText("Pay Over The Counter");
Button btnAccountToAccount = new Button(shell, SWT.NONE);
btnAccountToAccount.addSelectionListener(new SelectionAdapter() {
    @Override
    public void widgetSelected(SelectionEvent arg0) {
        try {
            AccountToAccount window = new AccountToAccount();
            window.open();
        } catch (Exception e) {
            e.printStackTrace(); }}});
btnAccountToAccount.setText("Account To Account");
btnAccountToAccount.setBounds(89, 100, 140, 33);
Button btnCashWithdrawal = new Button(shell, SWT.NONE);
btnCashWithdrawal.addSelectionListener(new SelectionAdapter() {
    @Override
    public void widgetSelected(SelectionEvent arg0) {
        try {
            CashWithdrawal window = new CashWithdrawal();
```

```java
            window.open();
        } catch (Exception e) {
            e.printStackTrace();   }}});
btnCashWithdrawal.setText("Cash Withdrawal");
btnCashWithdrawal.setBounds(89, 160, 140, 33);
Button btnCashDeposit = new Button(shell, SWT.NONE);
btnCashDeposit.addSelectionListener(new SelectionAdapter() {
    @Override
    public void widgetSelected(SelectionEvent arg0) {
        try {
            CashDeposit window = new CashDeposit();
            window.open();
        } catch (Exception e) {
            e.printStackTrace();
        }}});
btnCashDeposit.setText("Cash Deposit");
btnCashDeposit.setBounds(89, 222, 140, 33);
Button btnCheckBalance = new Button(shell, SWT.NONE);
btnCheckBalance.addSelectionListener(new SelectionAdapter() {
    @Override
    public void widgetSelected(SelectionEvent arg0) {
```

```java
            AddCustomer window = new AddCustomer();

            window.open();}});

btnCheckBalance.setText("Add Customer");

btnCheckBalance.setBounds(295, 43, 140, 33);

Button btnVerifyAccount = new Button(shell, SWT.NONE);

btnVerifyAccount.addSelectionListener(new SelectionAdapter() {

    @Override

    public void widgetSelected(SelectionEvent arg0) {

        VerifyAccount window = new VerifyAccount();

        window.open();}});

btnVerifyAccount.setText("Verify Account");

btnVerifyAccount.setBounds(295, 100, 140, 33);

Button btnVerifyPin = new Button(shell, SWT.NONE);

btnVerifyPin.addSelectionListener(new SelectionAdapter() {

    @Override

    public void widgetSelected(SelectionEvent arg0) {

        VerifyPIN window = new VerifyPIN();

        window.open();}});

btnVerifyPin.setText("Verify PIN");

btnVerifyPin.setBounds(295, 160, 140, 33);

Button btnSettings = new Button(shell, SWT.NONE);
```

```java
        btnSettings.addSelectionListener(new SelectionAdapter() {
            @Override
            public void widgetSelected(SelectionEvent arg0) {
                CancelTransaction window = new CancelTransaction();
                window.open();}});
        btnSettings.setText("Cancel Transaction");
        btnSettings.setBounds(295, 222, 140, 33);}}
```

### Pay Over The Counter

```java
package localhost.axis.JunoS_jws;

import java.rmi.RemoteException;

import javax.xml.rpc.ServiceException;

import org.eclipse.swt.widgets.Display;

import org.eclipse.swt.widgets.Shell;

import org.eclipse.swt.widgets.Text;

import org.eclipse.swt.SWT;

import org.eclipse.swt.widgets.Label;

import org.eclipse.swt.widgets.Button;

import org.eclipse.swt.events.SelectionAdapter;

import org.eclipse.swt.events.SelectionEvent;

import org.eclipse.swt.widgets.Canvas;

import org.eclipse.wb.swt.SWTResourceManager;
```

```java
public class PayOverTheCounter {

protected Shell shell;

    private Text text;

    private Text text_1;

    private Text text_2;

    public String x;

    public String y;

    public String z;

    /**
     * Launch the application.
     * @param args
     */
    public static void main(String[] args) {

        try {

            PayOverTheCounter window = new PayOverTheCounter();

            window.open();

        } catch (Exception e) {e.printStackTrace();}}

    /**
     * Open the window.
     */
    public void open() {
```

```java
            Display display = Display.getDefault();

            createContents();

            shell.open();

            shell.layout();

            while (!shell.isDisposed()) {

                    if (!display.readAndDispatch()) {

                            display.sleep();}}}
    /**
     * Create contents of the window.
     */
    protected void createContents() {

            shell = new Shell();

            shell.setSize(450, 300);

            shell.setText("Pay Over The Counter");

            text = new Text(shell, SWT.BORDER);

            text.setBounds(231, 71, 76, 21);

            text_1 = new Text(shell, SWT.PASSWORD);

            text_1.setBounds(231, 102, 76, 21);

        text_2 = new Text(shell, SWT.BORDER);

            text_2.setBounds(231, 138, 76, 21);

            Button btnOk = new Button(shell, SWT.NONE);
```

```java
btnOk.addSelectionListener(new SelectionAdapter() {
    @Override
    public void widgetSelected(SelectionEvent arg0) {
        JunoSService ss=new JunoSServiceLocator();
        try {
            JunoS s = (JunoS) ss.getJunoS();
            s.connection();
            x=text.getText();
            y=text_1.getText();
            z=text_2.getText();
            int success = s.payoverthecounter(x, y, z);
            if (success==1)
            {
                SuccessMessage window = new SuccessMessage();
                window.open();}
            else
            {
                ErrorMessage window = new ErrorMessage();
                window.open();
            }
            //    System.out.println(x);
```

```java
            } catch (ServiceException | RemoteException e) {
                // TODO Auto-generated catch block
                e.printStackTrace();}}});
        btnOk.setBounds(258, 210, 75, 25);
        btnOk.setText("OK");
        Label lblEnterAmount = new Label(shell, SWT.NONE);
        lblEnterAmount.setBounds(57, 74, 86, 15);
        lblEnterAmount.setText("Enter Amount:");
        Label lblEnterPin = new Label(shell, SWT.NONE);
        lblEnterPin.setBounds(57, 108, 55, 15);
        lblEnterPin.setText("Enter PIN:");
        Label lblEnterAccount = new Label(shell, SWT.NONE);
        lblEnterAccount.setBounds(57, 141, 86, 15);
        lblEnterAccount.setText("Enter Account:");
        Button btnCancel = new Button(shell, SWT.NONE);
        btnCancel.addSelectionListener(new SelectionAdapter() {
            @Override
            public void widgetSelected(SelectionEvent arg0) {
                JunoSService ss=new JunoSServiceLocator();
                shell.dispose();}});
        btnCancel.setBounds(339, 210, 75, 25);
```

```
            btnCancel.setText("Cancel");

            Label lblCancelTransaction = new Label(shell, SWT.NONE);

            lblCancelTransaction.setFont(SWTResourceManager.getFont("Segoe UI", 11, SWT.BOLD));

      lblCancelTransaction.setForeground(SWTResourceManager.getColor(SWT.COLOR_DARK_RED));

            lblCancelTransaction.setBounds(145, 22, 151, 21);

            lblCancelTransaction.setText("Pay Over The Counter");}}
```

## Account To Account

```java
package localhost.axis.JunoS_jws;

import java.rmi.RemoteException;

import javax.xml.rpc.ServiceException;

import org.eclipse.swt.widgets.Display;

import org.eclipse.swt.widgets.Shell;

import org.eclipse.swt.widgets.Text;

import org.eclipse.swt.SWT;

import org.eclipse.swt.widgets.Label;

import org.eclipse.swt.widgets.Button;

import org.eclipse.swt.events.SelectionAdapter;

import org.eclipse.swt.events.SelectionEvent;

import org.eclipse.swt.widgets.Canvas;
```

```java
import org.eclipse.wb.swt.SWTResourceManager;

public class AccountToAccount {

    protected Shell shell;
    private Text text;
    private Text text_1;
    private Text text_2;
    public String x;
    public String y;
    public String z;
    /**
     * Launch the application.
     * @param args
     */
    public static void main(String[] args) {
        try {
            AccountToAccount window = new AccountToAccount();
            window.open();
        } catch (Exception e) {
            e.printStackTrace();}}/**
    * Open the window.
    */
```

```java
public void open() {

    Display display = Display.getDefault();

    createContents();

    shell.open();

    shell.layout();

    while (!shell.isDisposed()) {

        if (!display.readAndDispatch()) {

            display.sleep();}}}
```
/**

 * Create contents of the window.

 */

```java
protected void createContents() {

    shell = new Shell();

    shell.setSize(450, 300);

    shell.setText("Account To Account");

    text = new Text(shell, SWT.BORDER);

    text.setBounds(231, 71, 76, 21);

    text_1 = new Text(shell, SWT.PASSWORD);

    text_1.setBounds(231, 102, 76, 21);

    text_2 = new Text(shell, SWT.BORDER);

    text_2.setBounds(231, 138, 76, 21);
```

```java
Button btnOk = new Button(shell, SWT.NONE);
btnOk.addSelectionListener(new SelectionAdapter() {
    @Override
    public void widgetSelected(SelectionEvent arg0) {
        JunoSService ss=new JunoSServiceLocator();
        try {
            JunoS s = (JunoS) ss.getJunoS();
            s.connection();
            x=text.getText();
            y=text_1.getText();
            z=text_2.getText();
            int result =    s.a2A(x, y, z);
            if (result==1)
            {
                SuccessMessage window = new SuccessMessage();
                window.open();
            }
            else
            {ErrorMessage window = new ErrorMessage();
                window.open();
            }} catch (ServiceException | RemoteException e) {
```

```java
				// TODO Auto-generated catch block
				e.printStackTrace();}}});
btnOk.setBounds(258, 210, 75, 25);
btnOk.setText("OK");

Label lblEnterAmount = new Label(shell, SWT.NONE);
lblEnterAmount.setBounds(57, 74, 86, 15);
lblEnterAmount.setText("Enter Amount:");

Label lblEnterPin = new Label(shell, SWT.NONE);
lblEnterPin.setBounds(57, 108, 55, 15);
lblEnterPin.setText("Enter PIN:");

Label lblEnterAccount = new Label(shell, SWT.NONE);
lblEnterAccount.setBounds(57, 141, 86, 15);
lblEnterAccount.setText("Enter Account:");

Button btnCancel = new Button(shell, SWT.NONE);
btnCancel.addSelectionListener(new SelectionAdapter() {
	@Override
	public void widgetSelected(SelectionEvent arg0) {
		shell.close();}});
btnCancel.setBounds(339, 210, 75, 25);
btnCancel.setText("Cancel");

Label lblCancelTransaction = new Label(shell, SWT.NONE);
```

```java
            lblCancelTransaction.setFont(SWTResourceManager.getFont("Segoe UI", 11, SWT.BOLD));
            lblCancelTransaction.setForeground(SWTResourceManager.getColor(SWT.COLOR_DARK_RED));
            lblCancelTransaction.setBounds(145, 22, 151, 21);
            lblCancelTransaction.setText("Account To Account");}}
```

## Add Customer

```java
package localhost.axis.JunoS_jws;

import java.rmi.RemoteException;

import javax.xml.rpc.ServiceException;

import org.eclipse.swt.widgets.Display;

import org.eclipse.swt.widgets.Shell;

import org.eclipse.swt.widgets.Label;

import org.eclipse.swt.SWT;

import org.eclipse.wb.swt.SWTResourceManager;

import org.eclipse.swt.widgets.Text;

import org.eclipse.swt.widgets.Button;

import org.eclipse.swt.events.SelectionAdapter;

import org.eclipse.swt.events.SelectionEvent;

public class AddCustomer {
protected Shell shell;
    private Text text;
```

```java
        private Text text_1;

        private Text text_2;

        private Text text_3;

        private Text text_4;

        private Text text_5;

        private Text text_6;

        private Text text_7;

        private Text text_8;

        private Text text_9;

        private Text text_10;

        String a,b,c,d,e,f,g,h,i,j,k;
    /**
         * Launch the application.
         * @param args
         */
        public static void main(String[] args) {
            try {
                AddCustomer window = new AddCustomer();
                window.open();
            } catch (Exception e) {
                e.printStackTrace();}}/**
```

```java
     * Open the window.
     */
    public void open() {
        Display display = Display.getDefault();
        createContents();
        shell.open();
        shell.layout();
        while (!shell.isDisposed()) {
            if (!display.readAndDispatch()) {
                display.sleep();
            }}}
    protected void createContents() {
        shell = new Shell();
        shell.setSize(439, 549);
        shell.setText("Add Customer");
        Label lblAddCustomer = new Label(shell, SWT.NONE);
        lblAddCustomer.setFont(SWTResourceManager.getFont("Segoe UI", 12, SWT.BOLD));
        lblAddCustomer.setForeground(SWTResourceManager.getColor(SWT.COLOR_DARK_RED));
        lblAddCustomer.setBounds(146, 25, 118, 26);
        lblAddCustomer.setText("Add Customer");
```

```java
Label lblCustomerId = new Label(shell, SWT.NONE);

lblCustomerId.setBounds(54, 75, 77, 15);

lblCustomerId.setText("Customer ID:");

text = new Text(shell, SWT.BORDER);

text.setBounds(226, 72, 97, 21);

Label lblNewLabel = new Label(shell, SWT.NONE);

lblNewLabel.setBounds(54, 110, 97, 15);

lblNewLabel.setText("Customer Name:");

text_1 = new Text(shell, SWT.BORDER);

text_1.setBounds(226, 107, 97, 21);

Label lblCustomerPhone = new Label(shell, SWT.NONE);

lblCustomerPhone.setBounds(54, 145, 97, 15);

lblCustomerPhone.setText("Customer Phone:");

Label lblCustomerOffice = new Label(shell, SWT.NONE);

lblCustomerOffice.setBounds(54, 181, 97, 15);

lblCustomerOffice.setText("Customer Office:");

Label lblCustomerFloor = new Label(shell, SWT.NONE);

lblCustomerFloor.setBounds(54, 220, 97, 15);

lblCustomerFloor.setText("Customer Floor:");

Label lblCustomerEmail = new Label(shell, SWT.NONE);

lblCustomerEmail.setBounds(54, 257, 97, 15);
```

```java
lblCustomerEmail.setText("Customer Email:");

Label lblCustomerAccount = new Label(shell, SWT.NONE);

lblCustomerAccount.setBounds(54, 291, 112, 15);

lblCustomerAccount.setText("Customer Account:");

Label lblNewLabel_1 = new Label(shell, SWT.NONE);

lblNewLabel_1.setBounds(54, 327, 131, 15);

lblNewLabel_1.setText("Customer Balance:");

Label lblCustomerUsername = new Label(shell, SWT.NONE);

lblCustomerUsername.setBounds(54, 365, 131, 15);

lblCustomerUsername.setText("Customer Username:");

Label lblCustomerPassword = new Label(shell, SWT.NONE);

lblCustomerPassword.setBounds(54, 404, 112, 15);

lblCustomerPassword.setText("Customer Password:");

Label lblCustomerPin = new Label(shell, SWT.NONE);

lblCustomerPin.setBounds(54, 438, 97, 15);

lblCustomerPin.setText("Customer PIN:");

text_2 = new Text(shell, SWT.BORDER);

text_2.setBounds(226, 145, 97, 21);

text_3 = new Text(shell, SWT.BORDER);

text_3.setBounds(226, 181, 97, 21);

text_4 = new Text(shell, SWT.BORDER);
```

```java
            text_4.setBounds(226, 217, 97, 21);
            
            text_5 = new Text(shell, SWT.BORDER);
            
            text_5.setBounds(226, 257, 97, 21);
            
            text_6 = new Text(shell, SWT.BORDER);
            
            text_6.setBounds(226, 291, 97, 21);
            
            text_7 = new Text(shell, SWT.BORDER);
            
            text_7.setBounds(226, 327, 97, 21);
            
    text_8 = new Text(shell, SWT.BORDER);
            
            text_8.setBounds(226, 365, 97, 21);
            
            text_9 = new Text(shell, SWT.BORDER);
            
            text_9.setBounds(227, 398, 96, 21);
            
            text_10 = new Text(shell, SWT.BORDER);
            
            text_10.setBounds(226, 435, 97, 21);
            
            Button btnNewButton = new Button(shell, SWT.NONE);
            
            btnNewButton.addSelectionListener(new SelectionAdapter() {
                
                @Override
                
                public void widgetSelected(SelectionEvent arg0) {
                    
                        JunoSService ss=new JunoSServiceLocator();
                        
                            try {
                            
                        JunoS s = (JunoS) ss.getJunoS();
                        
                            s.connection();
```

```java
a=text.getText();

b=text_1.getText();

c=text_2.getText();

d=text_3.getText();

e=text_4.getText();

f=text_5.getText();

g=text_6.getText();

h=text_7.getText();

i=text_8.getText();

j=text_9.getText();

k=text_10.getText();

int x =0;

x = s.addcustomer(a,b,c,d,e,f,g,h,i,j,k);

if (x == 1)
{
        SuccessMessage window = new SuccessMessage();

        window.open();
}
else if (x==2)
{ErrorMessage window = new ErrorMessage();

        window.open();}
```

```java
//        System.out.println(x);
        } catch (ServiceException | RemoteException e) {
            // TODO Auto-generated catch block
            e.printStackTrace();
        }}});
    btnNewButton.setBounds(238, 476, 75, 25);
    btnNewButton.setText("OK");
    Button btnNewButton_1 = new Button(shell, SWT.NONE);
    btnNewButton_1.addSelectionListener(new SelectionAdapter() {
        @Override
        public void widgetSelected(SelectionEvent arg0) {
            shell.close();}});
    btnNewButton_1.setBounds(326, 476, 75, 25);
    btnNewButton_1.setText("Cancel");}}
```

### Cash Deposit

```java
package localhost.axis.JunoS_jws;

import java.rmi.RemoteException;

import javax.xml.rpc.ServiceException;

import org.eclipse.swt.widgets.Display;

import org.eclipse.swt.widgets.Shell;

import org.eclipse.swt.widgets.Text;
```

```java
import org.eclipse.swt.SWT;

import org.eclipse.swt.widgets.Label;

import org.eclipse.swt.widgets.Button;

import org.eclipse.swt.events.SelectionAdapter;

import org.eclipse.swt.events.SelectionEvent;

import org.eclipse.swt.widgets.Canvas;

import org.eclipse.wb.swt.SWTResourceManager;

public class CashDeposit {

protected Shell shell;

    private Text text;

    private Text text_1;

    private Text text_2;

    public String x;

    public String y;

    public String z;

    /**
     * Launch the application.
     * @param args
     */
    public static void main(String[] args) {

        try {
```

```java
            CashDeposit window = new CashDeposit();
            window.open();
        } catch (Exception e) {
            e.printStackTrace();}}
/**
 * Open the window.
 */
    public void open() {
        Display display = Display.getDefault();
        createContents();
        shell.open();
        shell.layout();
        while (!shell.isDisposed()) {
            if (!display.readAndDispatch()) {
                display.sleep();}}}
/**
 * Create contents of the window.
 */
    protected void createContents() {
        shell = new Shell();
        shell.setSize(450, 300);
        shell.setText("Cash Deposit");
```

```java
        text = new Text(shell, SWT.BORDER);

        text.setBounds(231, 71, 76, 21);

    text_1 = new Text(shell, SWT.PASSWORD);

        text_1.setBounds(231, 102, 76, 21);

        text_2 = new Text(shell, SWT.BORDER);

        text_2.setBounds(231, 138, 76, 21);

        Button btnOk = new Button(shell, SWT.NONE);

        btnOk.addSelectionListener(new SelectionAdapter() {

            @Override

            public void widgetSelected(SelectionEvent arg0) {

                        JunoSService ss=new JunoSServiceLocator();

            try {

                        JunoS s = (JunoS) ss.getJunoS();

                            s.connection();

                            x=text.getText();

                            y=text_1.getText();

                            z=text_2.getText();

                        int success =     s.cashdeposit(x, y, z);

                        if (success==1)

                        {

                                SuccessMessage window = new SuccessMessage();
```

118 | P a g e

```java
                    window.open();
                }
                else
                {
                    ErrorMessage window = new ErrorMessage();
                    window.open();
                }
                //     System.out.println(x);
            } catch (ServiceException | RemoteException e) {
                // TODO Auto-generated catch block
                e.printStackTrace();}}});
        btnOk.setBounds(258, 210, 75, 25);
        btnOk.setText("OK");
        Label lblEnterAmount = new Label(shell, SWT.NONE);
        lblEnterAmount.setBounds(57, 74, 86, 15);
        lblEnterAmount.setText("Enter Amount:");
        Label lblEnterPin = new Label(shell, SWT.NONE);
        lblEnterPin.setBounds(57, 108, 55, 15);
        lblEnterPin.setText("Enter PIN:");
        Label lblEnterAccount = new Label(shell, SWT.NONE);
```

```java
        lblEnterAccount.setBounds(57, 141, 86, 15);

        lblEnterAccount.setText("Enter Account:");

        Button btnCancel = new Button(shell, SWT.NONE);

        btnCancel.addSelectionListener(new SelectionAdapter() {

            @Override

            public void widgetSelected(SelectionEvent arg0) {

                shell.close();}});

        btnCancel.setBounds(339, 210, 75, 25);

        btnCancel.setText("Cancel");

        Label lblCancelTransaction = new Label(shell, SWT.NONE);

        lblCancelTransaction.setFont(SWTResourceManager.getFont("Segoe UI", 11, SWT.BOLD));

        lblCancelTransaction.setForeground(SWTResourceManager.getColor(SWT.COLOR_DARK_RED));

        lblCancelTransaction.setBounds(145, 22, 151, 21);

        lblCancelTransaction.setText("Cash Deposit");}}
```

## Cash Withdrawal

```java
package localhost.axis.JunoS_jws;

import java.rmi.RemoteException;

import javax.xml.rpc.ServiceException;

import org.eclipse.swt.widgets.Display;

import org.eclipse.swt.widgets.Shell;
```

```java
import org.eclipse.swt.widgets.Text;

import org.eclipse.swt.SWT;

import org.eclipse.swt.widgets.Label;

import org.eclipse.swt.widgets.Button;

import org.eclipse.swt.events.SelectionAdapter;

import org.eclipse.swt.events.SelectionEvent;

import org.eclipse.swt.widgets.Canvas;

import org.eclipse.wb.swt.SWTResourceManager;

public class CashWithdrawal {

    protected Shell shell;

    private Text text;

    private Text text_1;

    private Text text_2;

    public String x;

    public String y;

    public String z;

    /**
     * Launch the application.
     * @param args
     */
    public static void main(String[] args) {
```

```java
        try {
            CashWithdrawal window = new CashWithdrawal();
            window.open();
        } catch (Exception e) {
            e.printStackTrace();}}
    /**
     * Open the window.
     */
    public void open() {
        Display display = Display.getDefault();
        createContents();
        shell.open();
        shell.layout();
        while (!shell.isDisposed()) {
            if (!display.readAndDispatch()) {
                display.sleep();}}     /**
     * Create contents of the window.
     */
    protected void createContents() {
        shell = new Shell();
        shell.setSize(450, 300);
```

```java
shell.setText("Cash Withdrawal");

text = new Text(shell, SWT.BORDER);

text.setBounds(231, 71, 76, 21);

text_1 = new Text(shell, SWT.PASSWORD);

text_1.setBounds(231, 102, 76, 21);

text_2 = new Text(shell, SWT.BORDER);

text_2.setBounds(231, 138, 76, 21);

Button btnOk = new Button(shell, SWT.NONE);

btnOk.addSelectionListener(new SelectionAdapter() {

    @Override
    public void widgetSelected(SelectionEvent arg0) {

        JunoSService ss=new JunoSServiceLocator();

            try {

            JunoS s = (JunoS) ss.getJunoS();

                s.connection();

                x=text.getText();

                y=text_1.getText();

                z=text_2.getText();

                int success = s.cashwithdrawal(x, y, z);

                if (success ==1)

                {
```

```java
                        SuccessMessage window = new SuccessMessage();
                            window.open();}
                    else
                    {
                    ErrorMessage window = new ErrorMessage();
                        window.open();}
            //      System.out.println(x);
                } catch (ServiceException | RemoteException e) {
                // TODO Auto-generated catch block
                e.printStackTrace();}}});
        btnOk.setBounds(258, 210, 75, 25);
        btnOk.setText("OK");
        Label lblEnterAmount = new Label(shell, SWT.NONE);
        lblEnterAmount.setBounds(57, 74, 86, 15);
        lblEnterAmount.setText("Enter Amount:");
        Label lblEnterPin = new Label(shell, SWT.NONE);
        lblEnterPin.setBounds(57, 108, 55, 15);
        lblEnterPin.setText("Enter PIN:");
        Label lblEnterAccount = new Label(shell, SWT.NONE);
        lblEnterAccount.setBounds(57, 141, 86, 15);
        lblEnterAccount.setText("Enter Account:");
```

```java
            Button btnCancel = new Button(shell, SWT.NONE);
            btnCancel.addSelectionListener(new SelectionAdapter() {
                @Override
                public void widgetSelected(SelectionEvent arg0) {
                    shell.close();   }});
            btnCancel.setBounds(339, 210, 75, 25);
            btnCancel.setText("Cancel");
            Label lblCancelTransaction = new Label(shell, SWT.NONE);
            lblCancelTransaction.setFont(SWTResourceManager.getFont("Segoe UI", 11, SWT.BOLD));
            lblCancelTransaction.setForeground(SWTResourceManager.getColor(SWT.COLOR_DARK_RED));
            lblCancelTransaction.setBounds(145, 22, 151, 21);
            lblCancelTransaction.setText("Cash Withdrawal");}}
```

## Cancel Transaction

```java
package localhost.axis.JunoS_jws;

import java.rmi.RemoteException;

import javax.xml.rpc.ServiceException;

import org.eclipse.swt.widgets.Display;

import org.eclipse.swt.widgets.Shell;

import org.eclipse.swt.widgets.Text;
```

```java
import org.eclipse.swt.SWT;

import org.eclipse.swt.widgets.Label;

import org.eclipse.swt.widgets.Button;

import org.eclipse.swt.events.SelectionAdapter;

import org.eclipse.swt.events.SelectionEvent;

import org.eclipse.swt.widgets.Canvas;

import org.eclipse.wb.swt.SWTResourceManager;

public class CancelTransaction {

    protected Shell shell;

    private Text text;

    private Text text_1;

    private Text text_2;

    public String x;

    public String y;

    public String z;

    /**
     * Launch the application.
     * @param args
     */
    public static void main(String[] args) {

        try {
```

```java
                CancelTransaction window = new CancelTransaction();
                window.open();
        } catch (Exception e) {
            e.printStackTrace();}}
/**    * Open the window.
     */
    public void open() {
        Display display = Display.getDefault();
        createContents();
        shell.open();
        shell.layout();
        while (!shell.isDisposed()) {
            if (!display.readAndDispatch()) {
                display.sleep();       }}}
/**
     * Create contents of the window.
     */
    protected void createContents() {
        shell = new Shell();
        shell.setSize(450, 300);
        shell.setText("Cancel Transaction");
```

```java
text = new Text(shell, SWT.BORDER);
text.setBounds(231, 71, 76, 21);
text_1 = new Text(shell, SWT.PASSWORD);
text_1.setBounds(231, 102, 76, 21);
text_2 = new Text(shell, SWT.BORDER);
text_2.setBounds(231, 138, 76, 21);
Button btnOk = new Button(shell, SWT.NONE);
btnOk.addSelectionListener(new SelectionAdapter() {
    @Override
    public void widgetSelected(SelectionEvent arg0) {
        JunoSService ss=new JunoSServiceLocator();
            try {
            JunoS s = (JunoS) ss.getJunoS();
                s.connection();
                x=text.getText();
                y=text_1.getText();
                z=text_2.getText();
            int success = s.canceltransaction(x, y, z);
            if(success == 1)
            {
                SuccessMessage window = new SuccessMessage();
```

```java
                    window.open();}
        else
        {
            ErrorMessage window = new ErrorMessage();
            window.open();}
//      System.out.println(x);
        } catch (ServiceException | RemoteException e) {
            // TODO Auto-generated catch block
            e.printStackTrace();}}});
btnOk.setBounds(258, 210, 75, 25);

btnOk.setText("OK");

Label lblEnterAmount = new Label(shell, SWT.NONE);

lblEnterAmount.setBounds(57, 74, 86, 15);

lblEnterAmount.setText("Enter Amount:");

Label lblEnterPin = new Label(shell, SWT.NONE);

lblEnterPin.setBounds(57, 108, 55, 15);

lblEnterPin.setText("Enter PIN:");

Label lblEnterAccount = new Label(shell, SWT.NONE);

lblEnterAccount.setBounds(57, 141, 86, 15);

lblEnterAccount.setText("Enter Account:");

Button btnCancel = new Button(shell, SWT.NONE);
```

```java
            btnCancel.addSelectionListener(new SelectionAdapter() {
                @Override
                public void widgetSelected(SelectionEvent arg0) {
                    shell.close();}});
        btnCancel.setBounds(339, 210, 75, 25);
        btnCancel.setText("Cancel");
        Label lblCancelTransaction = new Label(shell, SWT.NONE);
        lblCancelTransaction.setFont(SWTResourceManager.getFont("Segoe UI", 11, SWT.BOLD));

        lblCancelTransaction.setForeground(SWTResourceManager.getColor(SWT.COLOR_DARK_RED));
        lblCancelTransaction.setBounds(145, 22, 151, 21);
        lblCancelTransaction.setText("Cancel Transaction");
    }
}
```

## Success Message

```java
package localhost.axis.JunoS_jws;

import org.eclipse.swt.widgets.Display;

import org.eclipse.swt.widgets.Shell;

import org.eclipse.swt.widgets.Label;

import org.eclipse.swt.SWT;
```

```java
import org.eclipse.wb.swt.SWTResourceManager;

public class SuccessMessage {
    protected Shell shell;
    /**
     * Launch the application.
     * @param args
     */
    public static void main(String[] args) {
        try {
            SuccessMessage window = new SuccessMessage();
            window.open();
        } catch (Exception e) {
            e.printStackTrace();}}
    /**
     * Open the window.
     */
    public void open() {
        Display display = Display.getDefault();
        createContents();
        shell.open();
```

```
                shell.layout();

                while (!shell.isDisposed()) {

                        if (!display.readAndDispatch()) {

                                display.sleep();}}}
```
/**

   * Create contents of the window.

   */

```
    protected void createContents() {

            shell = new Shell();

            shell.setSize(268, 162);

            shell.setText("Proceed");

            Label lblErrorTryAgain = new Label(shell, SWT.NONE);

            lblErrorTryAgain.setFont(SWTResourceManager.getFont("Segoe UI", 12, SWT.NORMAL));

            lblErrorTryAgain.setBounds(20, 10, 68, 21);

            lblErrorTryAgain.setText("Message: ");

            Label lblNewLabel = new Label(shell, SWT.NONE);

            lblNewLabel.setBounds(20, 47, 222, 15);

            lblNewLabel.setText("Executed.");    }}
```

## Error Message

package localhost.axis.JunoS_jws;

132 | P a g e

```java
import org.eclipse.swt.widgets.Display;

import org.eclipse.swt.widgets.Shell;

import org.eclipse.swt.widgets.Label;

import org.eclipse.swt.SWT;

import org.eclipse.wb.swt.SWTResourceManager;

public class ErrorMessage {

protected Shell shell;

    /**
     * Launch the application.
     * @param args
     */
    public static void main(String[] args) {

        try {

            ErrorMessage window = new ErrorMessage();

            window.open();

        } catch (Exception e) {

            e.printStackTrace();}}

    /**
     * Open the window.
     */
    public void open() {
```

```java
            Display display = Display.getDefault();
            createContents();
            shell.open();
            shell.layout();
            while (!shell.isDisposed()) {
                if (!display.readAndDispatch()) {
                    display.sleep();
                }}}
    /**
     * Create contents of the window.
     */
    protected void createContents() {
        shell = new Shell();
        shell.setSize(268, 162);
        shell.setText("Error Message");
        Label lblErrorTryAgain = new Label(shell, SWT.NONE);
        lblErrorTryAgain.setFont(SWTResourceManager.getFont("Segoe UI", 12, SWT.NORMAL));
        lblErrorTryAgain.setBounds(20, 10, 58, 21);
        lblErrorTryAgain.setText("Error: ");
        Label lblNewLabel = new Label(shell, SWT.NONE);
```

```
        lblNewLabel.setBounds(20, 47, 222, 15);

        lblNewLabel.setText("Try Again Later.");

    }}
```

## Verify PIN

```java
package localhost.axis.JunoS_jws;

import java.rmi.RemoteException;

import javax.xml.rpc.ServiceException;

import org.eclipse.swt.widgets.Display;

import org.eclipse.swt.widgets.Shell;

import org.eclipse.swt.widgets.Text;

import org.eclipse.swt.SWT;

import org.eclipse.swt.widgets.Label;

import org.eclipse.swt.widgets.Button;

import org.eclipse.wb.swt.SWTResourceManager;

import org.eclipse.swt.events.SelectionAdapter;

import org.eclipse.swt.events.SelectionEvent;

public class VerifyPIN {

protected Shell shell;

    private Text text;

    public String x;
```

```java
String c = "Try Again";
    /**
     * Launch the application.
     * @param args
     */
    public static void main(String[] args) {
        try {
            VerifyPIN window = new VerifyPIN();
            window.open();
        } catch (Exception e) {
            e.printStackTrace();}}
    /**
     * Open the window.
     */
    public void open() {
        Display display = Display.getDefault();
        createContents();
        shell.open();
        shell.layout();
        while (!shell.isDisposed()) {
            if (!display.readAndDispatch()) {
```

```java
                    display.sleep();
        }}}
    /**
     * Create contents of the window.
     */
    protected void createContents() {
        shell = new Shell();
        shell.setSize(450, 300);
        shell.setText("PIN Verification");
        text = new Text(shell, SWT.PASSWORD);
        text.setBounds(233, 79, 99, 21);
        Label lblEnterAccount = new Label(shell, SWT.NONE);
        lblEnterAccount.setBounds(100, 79, 99, 21);
        lblEnterAccount.setText("Enter PIN:");
        Button btnOk = new Button(shell, SWT.NONE);
        btnOk.addSelectionListener(new SelectionAdapter() {
            @Override
            public void widgetSelected(SelectionEvent arg0) {
                JunoSService ss=new JunoSServiceLocator();
                try {
                    JunoS s = (JunoS) ss.getJunoS();
                    s.connection();
```

```java
                        x=text.getText();

                        //      s.verifypin(x);

                //      System.out.println(x);

                         int a = 0;

                         a = s.checkpin(x);

                s.searchData("select * from Account where idAccount = '"+x+"'");

        //      System.out.println(a);

                          if(a==2)

                           { // System.out.println(a);

                            ErrorMessage window = new ErrorMessage();

                                    window.open();}

                            else if(a==1)

                             {

                                    Verified window = new Verified();

                                    window.open();           }

                        } catch (ServiceException | RemoteException e) {

                                // TODO Auto-generated catch block

                                e.printStackTrace();

                        }}});

        btnOk.setBounds(243, 227, 75, 25);

        btnOk.setText("OK");
```

```java
            Button btnCancel = new Button(shell, SWT.NONE);
            btnCancel.addSelectionListener(new SelectionAdapter() {
                @Override
                public void widgetSelected(SelectionEvent arg0) {
                    shell.dispose();
                }});
            btnCancel.setBounds(329, 227, 75, 25);
            btnCancel.setText("Cancel");
            Label lblAccountVerification = new Label(shell, SWT.NONE);
            lblAccountVerification.setForeground(SWTResourceManager.getColor(SWT.COLOR_DARK_RED));
            lblAccountVerification.setFont(SWTResourceManager.getFont("Segoe UI", 11, SWT.BOLD));
            lblAccountVerification.setBounds(150, 21, 136, 21);
            lblAccountVerification.setText("PIN Verification");}}
```

## Verify Account

```java
package localhost.axis.JunoS_jws;

import java.rmi.RemoteException;

import javax.xml.rpc.ServiceException;

import org.eclipse.swt.widgets.Display;

import org.eclipse.swt.widgets.Shell;

import org.eclipse.swt.widgets.Text;
```

```java
import org.eclipse.swt.SWT;
import org.eclipse.swt.widgets.Label;
import org.eclipse.swt.widgets.Button;
import org.eclipse.wb.swt.SWTResourceManager;
import org.eclipse.swt.events.SelectionAdapter;
import org.eclipse.swt.events.SelectionEvent;
public class VerifyAccount {
    protected Shell shell;
    private Text text;
    public String x;
    /**
     * Launch the application.
     * @param args
     */
    public static void main(String[] args) {
        try {
            VerifyAccount window = new VerifyAccount();
            window.open();
        } catch (Exception e) {
            e.printStackTrace();}}
/**    * Open the window.
```

```java
 */
public void open() {
    Display display = Display.getDefault();
    createContents();
    shell.open();
    shell.layout();
    while (!shell.isDisposed()) {
        if (!display.readAndDispatch()) {
            display.sleep();
        }}}    /**
 * Create contents of the window.
 */
protected void createContents() {
    shell = new Shell();
    shell.setSize(450, 300);
    shell.setText("Account Verification");
    text = new Text(shell, SWT.BORDER);
    text.setBounds(243, 80, 99, 21);
    Label lblEnterAccount = new Label(shell, SWT.NONE);
    lblEnterAccount.setBounds(94, 83, 99, 21);
    lblEnterAccount.setText("Enter Account:");
```

```java
Button btnOk = new Button(shell, SWT.NONE);
btnOk.addSelectionListener(new SelectionAdapter() {
    @Override
    public void widgetSelected(SelectionEvent arg0) {
        JunoSService ss=new JunoSServiceLocator();
        try {
            JunoS s = (JunoS) ss.getJunoS();
            s.connection();
            x=text.getText();
            int a = 0;
            a = s.checkaccount(x);
            System.out.println(a);
            System.out.println(x);
            if(a==1)
            {   Verified window = new Verified();
                window.open();
                // System.out.println(a);
            }
            else if(a==2)
            { ErrorMessage window = new ErrorMessage();
                window.open();}
```

```java
                    //        System.out.println(x);
                } catch (ServiceException | RemoteException e) {
                    // TODO Auto-generated catch block
                    e.printStackTrace();
                }}});
        btnOk.setBounds(243, 227, 75, 25);
        btnOk.setText("OK");
        Button btnCancel = new Button(shell, SWT.NONE);
        btnCancel.addSelectionListener(new SelectionAdapter() {
            @Override
            public void widgetSelected(SelectionEvent arg0) {
                shell.dispose();
            }});
        btnCancel.setBounds(329, 227, 75, 25);
    btnCancel.setText("Cancel");
        Label lblAccountVerification = new Label(shell, SWT.NONE);
    lblAccountVerification.setForeground(SWTResourceManager.getColor(SWT.COLOR_DARK_RED));
        lblAccountVerification.setFont(SWTResourceManager.getFont("Segoe UI", 11, SWT.BOLD));
        lblAccountVerification.setBounds(142, 20, 151, 21);
        lblAccountVerification.setText("Account Verification");    }}
```

# Verification Message

```java
package localhost.axis.JunoS_jws;

import org.eclipse.swt.widgets.Display;

import org.eclipse.swt.widgets.Shell;

import org.eclipse.swt.widgets.Label;

import org.eclipse.swt.SWT;

import org.eclipse.wb.swt.SWTResourceManager;

public class Verified {

    protected Shell shell;
/**
     * Launch the application.
     * @param args
     */
    public static void main(String[] args) {
        try {
            SuccessMessage window = new SuccessMessage();
            window.open();
        } catch (Exception e) {
            e.printStackTrace();}}/**
    * Open the window.
    */
```

```java
public void open() {

    Display display = Display.getDefault();

    createContents();

    shell.open();

    shell.layout();

    while (!shell.isDisposed()) {

        if (!display.readAndDispatch()) {

            display.sleep();

        }}}
/**
 * Create contents of the window.
 * @wbp.parser.entryPoint
 */
protected void createContents() {

    shell = new Shell();

    shell.setSize(268, 162);

    shell.setText("Verification Message");

    Label lblErrorTryAgain = new Label(shell, SWT.NONE);

    lblErrorTryAgain.setFont(SWTResourceManager.getFont("Segoe UI", 12, SWT.NORMAL));

    lblErrorTryAgain.setBounds(20, 10, 68, 21);
```

```
lblErrorTryAgain.setText("Message: ");

Label lblNewLabel = new Label(shell, SWT.NONE);

lblNewLabel.setBounds(20, 47, 222, 15);

lblNewLabel.setText("Account Exists.");}}
```

# APPENDIX B

## Keypad Code

```
#include <Keypad.h>
const byte ROWS = 4; //four rows
const byte COLS = 4; //four columns
//define the cymbols on the buttons of the keypads
char hexaKeys[ROWS][COLS] = {
  {'1','2','3','A'},
  {'4','5','6','B'},
  {'7','8','9','C'},
  {'*','0','#','D'}};
byte rowPins[ROWS] = {22, 24, 26, 28}; //connect to the row pinouts of the keypad
byte colPins[COLS] = {30, 32, 34, 36}; //connect to the column pinouts of the keypad
//initialize an instance of class NewKeypad
Keypad customKeypad = Keypad( makeKeymap(hexaKeys), rowPins, colPins, ROWS, COLS);

void setup(){
  Serial.begin(9600);}
 void loop(){
 char customKey = customKeypad.getKey();
  if (customKey){
  Serial.println(customKey);}}
```

## Decoder Code

```
void blinkledfast(){
```

```
 for(int i = 0; i<40;i++){
 digitalWrite(13,HIGH);
 delay(50);
 digitalWrite(13,LOW);
 delay(50); }}
void blinkledslow(){
 for(int i = 0; i<10;i++){
 digitalWrite(13,HIGH);
 delay(500);
 digitalWrite(13,LOW);
 delay(500); }}
void setup(){
Serial.begin(9600);
Serial1.begin(9600);
Serial2.begin(9600);}
void loop(){
 char x, y;
 x=2;
 y=0;
Serial1.print(x);
Serial1.print(y);
while(!Serial.available());
while(Serial1.available()){
 //Serial.println("Communication started");
 Serial.println(Serial1.read());}
```

```
if(!Serial1.available()){
 Serial.println("not available"); }
delay(2000);}
```

## Ethernet Shield Code

```
#include <SPI.h>
#include <Ethernet.h>
byte ticker = 0;
// Enter a MAC address and IP address for your controller below.
// The IP address will be dependent on your local network:
byte mac[] = {
  0xDE, 0xAD, 0xBE, 0xEF, 0xFE, 0xED };
IPAddress ip(192,168,0, 200);
IPAddress sserver(192,168,0,100);
//char sserver[] = "www.google.com";
// Initialize the Ethernet server library
// with the IP address and port you want to use
// (port 80 is default for HTTP):
EthernetServer server(80);
EthernetClient sclient;

void setup() {
 // Open serial communications and wait for port to open:
 Serial.begin(9600);
  while (!Serial) {
  ;
```

```
 }
/*
  cli();
  //set timer1 interrupt at 1Hz
  TCCR1A = 0;// set entire TCCR1A register to 0
  TCCR1B = 0;// same for TCCR1B
  TCNT1  = 0;//initialize counter value to 0
  // set compare match register for 1hz increments
  OCR1A = 62496;// = (16*10^6) / (1*1024) - 1 (must be <65536)
  // turn on CTC mode
  TCCR1B |= (1 << WGM12);
  // Set CS12 and CS10 bits for 1024 prescaler
  TCCR1B |= (1 << CS12) | (1 << CS10);
  // enable timer compare interrupt
  TIMSK1 |= (1 << OCIE1A);
   sei();*/
  // start the Ethernet connection and the server:
  Ethernet.begin(mac, ip);
  server.begin();
  Serial.print("server is at ");
  Serial.println(Ethernet.localIP());
   // give the Ethernet shield a second to initialize:
  delay(1000);
  Serial.println("connecting...");}
//ISR(TIMER1_COMPA_vect){//timer1 interrupt 1Hz toggles pin 13 (LED)
```

//generates pulse wave of frequency 1Hz/2 = 0.5Hz (takes two cycles for full wave- toggle high then toggle low)

```
   // if you get a connection, report back via serial:
 /*if(ticker==3){
  if (sclient.connect(sserver, 80)) {
    Serial.println("connected");
    // Make a HTTP request:
    sclient.println("GET /post.php HTTP/1.1");
    sclient.println("Host: 192.168.1.6");
    sclient.println("Connection: close");
    sclient.println();
     Serial.println("REQUEST SENT");
  }
  else {
    // kf you didn't get a connection to the server:
    Serial.println("connection failed");
   }
  ticker=0;
  }
  else {
   ticker++;
  }*/
 // Serial.println("WHY??\n\n");
//}
void loop() {
```

```
  float z=0,b=0;
 for(int a=0;a<100;a++)
  {
   delay(25);
   float cValue = analogRead(A0);
   float vValue = analogRead(A1);
   float x= cValue-513;
   x=(x*5)/1024;
   //Serial.println(x);
   if(z<x)
    z=x;
   float y=vValue;
   if(b<y)
     b=y;
  }
  float c=z/0.085;     //c is current in amperes
  if(c<0.08)
    c=0;
  float v=b;
  v=(v*5)/1024;
  v=(50*v)+0;
  Serial.println("Current=");
  Serial.println(c);
  Serial.println("Voltage=");
  Serial.println(v);
```

```
    Serial.println("****************");

    Serial.println("Power=");

    Serial.println(c*v);

    Serial.println("****************");

    Serial.println(" ");

    Serial.println(" ");

     if (sclient.connect(sserver, 80)) {

     Serial.println("connected");

     // Make a HTTP request:

     sclient.print("GET /post.php?c=");

     sclient.print(c,DEC);

     sclient.print("&v=");

     sclient.print(v,DEC);

     sclient.print("&p=");

     sclient.print(c*v,DEC);

     sclient.println(" HTTP/1.1");

     sclient.println("Host: 192.168.1.6");

     sclient.println("Connection: close");

     sclient.println();

      Serial.println("REQUEST SENT");

        sclient.stop();

    }

    else {

     // kf you didn't get a connection to the server:

     Serial.println("connection failed");
```

```
  }
    //delay(1000);      // delay in between reads for stability
  // listen for incoming clients
  EthernetClient client = server.available();
    if (client) {
    Serial.println("new client");
    // an http request ends with a blank line
    boolean currentLineIsBlank = true;
    while (client.connected()) {
     if (client.available()) {
       char k = client.read();
       Serial.write(k);
       // if you've gotten to the end of the line (received a newline
       // character) and the line is blank, the http request has ended,
       // so you can send a reply
       if (k == '\n' && currentLineIsBlank) {
        // send a standard http response header
        client.println("HTTP/1.1 200 OK");
        client.println("Content-Type: text/html");
        client.println("Connection: close");
        client.println();
        client.println("<!DOCTYPE HTML>");
        client.println("<html>");
            // add a meta refresh tag, so the browser pulls again every 5 seconds:
        client.println("<meta http-equiv=\"refresh\" content=\"5\">");
```

```
      // output the value of each analog input pin
        client.print("CURRENT ");
        client.print(" is ");
        client.print(c);
        client.println("<br />");
   client.print("voltage ");
        client.print(" is ");
        client.print(v);
        client.println("<br />");
      client.print("power ");
        client.print(" is ");
        client.print(c*v);
        client.println("<br />");
          client.println("</html>");
       break;
     }
    if (k == '\n') {
      // you're starting a new line
     currentLineIsBlank = true;   }
    else if (k != '\r') {
      // you've gotten a character on the current line
     currentLineIsBlank = false;  }} }
// give the web browser time to receive the data
delay(1);
// close the connection:
```

```
client.stop();

Serial.println("client disonnected");}}
```